\documentclass[12pt]{article}

\usepackage[margin=1.0in]{geometry}

\usepackage{cite}

\usepackage{amssymb,latexsym}

\usepackage{epstopdf}

\usepackage{color}
\usepackage{xcolor}

\usepackage{amsmath,amsfonts}
\usepackage{empheq}

\usepackage{mathrsfs}

\usepackage{slashed}

\DeclareMathAlphabet{\mathpzc}{OT1}{pzc}{m}{it}

 \usepackage{mathtools}

\usepackage{feynmp}
\newcommand*{\Scale}[2][4]{\scalebox{#1}{$#2$}}%

\let\a=\alpha \let\b=\beta \let\g=\gamma \let\d=\delta \let\e=\epsilon
\let\z=\zeta \let\h=\eta \let\th=\theta  \let\k=\kappa
\let\l=\lambda \let\m=\mu \let\n=\nu \let\x=\xi \let\p=\pi 
\let\s=\sigma \let\t=\tau  \let\f=\phi  
 
\let\w=\omega      \let\G=\Gamma \let\D=\Delta \let\Th=\Theta \let\L=\Lambda
\let\X=\Xi  \let\S=\Sigma  \let\Y=\Psi
 
\let\la=\label  
  
\def\nn{\nonumber} \def\bd{\begin{document}} \def\ed{\end{document}}
\let\fr=\frac \let\bl=\bigl \let\br=\bigr
\let\Br=\Bigr \let\Bl=\Bigl
\let\bm=\bibitem
\let\na=\nabla
\def\tU{{\widetilde U}}
\let\pa=\partial \let\ov=\overline
\def\ie{{\it i.e.\ }}
\newcommand{\be}{\begin{equation}}
\newcommand{\ee}{\end{equation}}
\def\ba{\begin{array}}
\def\ea{\end{array}}
\def\ft#1#2{{\textstyle{{\scriptstyle #1}\over {\scriptstyle #2}}}}
\def\fft#1#2{{#1 \over #2}}
\def\cF#1#2{{ {\cal F}_{#1}^{(#2)} }}

\def\R{{\bf R}}
\def\sst#1{{\scriptscriptstyle #1}}
\def\oneone{\rlap 1\mkern4mu{\rm l}}
\def\e7{E_{7(+7)}}
\def\td{\tilde}
\def\wtd{\widetilde}
\def\im{{\rm i}}
\def\bog{Bogomol'nyi\ }
\newcommand{\ho}[1]{$\, ^{#1}$}
\newcommand{\hoch}[1]{$\, ^{#1}$}
\newcommand{\bea}{\begin{eqnarray}}
\newcommand{\eea}{\end{eqnarray}}

\newcommand{\ra}{\rightarrow}
\newcommand{\lra}{\longrightarrow}
\newcommand{\Lra}{\Leftrightarrow}
\newcommand{\ap}{\alpha^\prime}
\newcommand{\bp}{\tilde \beta^\prime}
\newcommand{\cB}{{\cal B}}
\newcommand{\cO}{{\cal O}}
\newcommand{\vecx}{\vec{x}}
\newcommand{\vecy}{\vec{y}}
\newcommand{\vecp}{\vec{p}}
\newcommand{\vecq}{\vec{q}}
\newcommand{\tr}{{\rm tr} }
\newcommand{\Tr}{{\rm Tr} }
\newcommand{\NP}{Nucl. Phys. }

\newcommand{\cL}{{\cal L}}
\newcommand{\cA}{{\cal A}}
\newcommand{\cT}{{\cal T}}
\newcommand{\cR}{{\cal R}}
\newcommand{\cD}{{\cal D}}
\newcommand{\cH}{{\cal H}}

\def\Cb{\bar{C}}

\def\sst#1{{\scriptscriptstyle #1}}
\def\0{{\sst{(0)}}}
\def\1{{\sst{(1)}}}
\def\2{{\sst{(2)}}}
\def\3{{\sst{(3)}}}
\def\4{{\sst{(4)}}}
\def\5{{\sst{(5)}}}
\def\6{{\sst{(6)}}}
\def\7{{\sst{(7)}}}
\def\8{{\sst{(8)}}}
\def\9{{\sst{(9)}}}
\def\p{{\sst{(p)}}}
\def\q{{\sst{(q)}}}
\def\ve{\varepsilon}
\def\vf{\varphi}
\def\vF{\varPhi}
\def\F{\Phi}
\def\wg{\wedge}

\def\e{\epsilon}
\def\barl{\bar{l}}

\def \bi{\bibitem}
\def \la {\label}

\def \l {\lambda}
\def\foot{\footnote}
\def \tl  {{\tilde \l}}
\def \sql {{\sqrt \l}}
\def \adss {$AdS_5 \times S^5$\ }
\newcommand{\rf}[1]{(\ref{#1})}
\def \ov {\over}

\def\th{\theta}
\def\Th{\Theta}
\def\vth{\vartheta}
\def\btheta{{\bar\theta}}
\def\ttheta{{{\tilde\theta}}}
\def\bttheta{{{\bar\ttheta}}}
\def\vth{\vartheta}

\def\ra{\rightarrow}
\def\N{\nabla}
\def\uM{\underline{M}}
\def\uA{\underline{A}}
\def\uN{\underline{N}}
\def\uP{\underline{P}}
\def\ua{\underline{a}}
\def\ub{\underline{b}}
\def\uc{\underline{c}}
\def\ud{\underline{d}}
\def\ue{\underline{e}}
\def\uf{\underline{f}}
\def\ui{\underline{i}}
\def\uj{\underline{j}}
\def\uk{\underline{k}}
\def\ul{\underline{l}}
\def\ual{\underline{\alpha}}
\def\ube{\underline{\beta}}
\def\um{\underline{m}}
\def\un{\underline{n}}
\def\up{\underline{p}}
\def\uq{\underline{q}}
\def\ur{\underline{r}}
\def\us{\underline{s}}
\def\umu{\underline{\mu}}
\def\unu{\underline{\nu}}
\def\ula{\underline{\l}}
\def\uka{\underline{\k}}
\def\usi{\underline{\s}}
\def\urh{\underline{\r}}
\def\cc{\circ}
\def\eqv{\equiv}

\def\ni{\noindent}

\def\Ep{E^{{}^{(+)}}}
\def\Em{E^{{}^{(-)}}}

\def\Mp{M^{{}^{(+)}}}
\def\Mm{M^{{}^{(-)}}}

\def \ha{{1\ov 2}}

\def\r{\rho}

\def\Y{{\rm Y}}
\def\X{{\rm X}}
\def\tY{\tilde{\rm Y}}
\def\tX{\tilde{\rm X}}
\def\dY{\dot{\rm Y}}
\def\dX{\dot{\rm X}}

\def \J {\mathcal{J}}
\def \del {\partial}

\def\dF{\dot{F}}
\def\dG{\dot{G}}
\def\df{\dot{f}}
\def \E {{\cal E}}
\def \S {{\cal S}}
\def \J {{\cal J}}

\def\ms{\mathcal{S}}
\def\mj{\mathcal{J}}
\def\soj{\fr{\ms}{\mj}}
\def \R {{\bf R}}
\def \om {\omega}
\def \bE {\bar E}
\def \x {{\cal X}}

\def \bi{\bibitem}
\def \la {\label}

\def \l {\lambda}
\def\foot{\footnote}
\def \tl  {{\tilde \l}}
\def \sql {{\sqrt \l}}
\def \adss {$AdS_5 \times S^5$\ }
\def \ov {\over}

\def \varpi {{\rm w}}

\def\thb{\bar{\theta}}
\def\Thb{\bar{\Theta}}
\def\barp{\bar{p}}
\def\barq{\bar{q}}
\def\barc{\bar{c}}
\def\bard{\bar{d}}
\def\bare{\bar{e}}

\def\thb{\bar{\theta}}
\def\Thb{\bar{\Theta}}
\def\mb{\bar{\m}}
\def\ab{\bar{\a}}
\def\zb{\bar{z}}
\def\psib{\bar{\psi}}
\def\barl{\bar{l}}
\def\barp{\bar{p}}
\def\barq{\bar{q}}
\def\barc{\bar{c}}
\def\bard{\bar{d}}
\def\baru{\bar{u}}

\def\e{\epsilon}
\def\wb{\bar{w}}
\def\lb{\bar{\l}}
\def\Jb{\bar{J}}
\def\Nb{\bar{N}}
\def\Zb{\bar{Z}}
\def\pab{\bar{\pa}}

\def\At{\tilde{A}}
\def\Bt{\tilde{B}}
\def\Ct{\tilde{C}}
\def\Dt{\tilde{D}}
\def\Et{\tilde{E}}
\def\Gt{\tilde{G}}
\def\Ht{\tilde{H}}
\def\Kt{\tilde{K}}
\def\Mt{\tilde{M}}
\def\Nt{\tilde{N}}
\def\Rt{\tilde{R}}
\def\at{\tilde{a}}
\def\bt{\tilde{b}}
\def\ct{\tilde{c}}
\def\dt{\tilde{d}}
\def\et{\tilde{e}}
\def\ft{\tilde{f}}
\def \ztt{\tilde{\z}}
\def \zetat{\tilde{\zeta}}
\def\htil{\tilde{h}}
\def\gt{\tilde{g}}
\def\nt{\tilde{n}}
\def\mut{\tilde{\mu}}
\def\nut{\tilde{\nu}}
\def\pht{\tilde{\f}}
\def\Phit{\tilde{\Phi}}
\def\vft{\tilde{\vf}}

\def\rht{\tilde{\rho}}

\def\asth{\hat{*}}
\def\phh{\hat{\phi}}

\def\bA{{\bf A}}

\def\ola{\overleftarrow}
\def\ora{\overrightarrow}
\def\alt{\tilde{\a}}

\def\eh{\hat{e}}
\def\eph{\hat{\e}}
\def\ph{\hat{p}}
\def\alh{\hat{\a}}
\def\beh{\hat{\b}}
\def\gah{\hat{\g}}
\def\Fh{\hat{F}}
\def\muh{\hat{\m}}
\def\nuh{\hat{\n}}
\def\thh{\hat{\th}}
\def\rhh{\hat{\r}}
\def\dh{\hat{d}}
\def\ih{\hat{i}}
\def\jh{\hat{j}}
\def\hh{\hat{h}}
\def\nh{\hat{n}}
\def\gh{\hat{g}}
\def\kh{\hat{k}}
\def\deh{\hat{\d}}
\def\wh{\hat{w}}
\def\lah{\hat{\l}}
\def\Ah{\hat{A}}
\def\Gh{\hat{G}}
\def\Kh{\hat{K}}
\def\Nh{\hat{N}}
\def\Rh{\hat{R}}
\def\Ch{\hat{C}}
\def\Omh{\hat{\Omega}}

\def\xh{\hat{x}}

\def\ps{\rlap{\, /}\;\,p }
\def\ks{\rlap{\, /}\;\,k }

\def\gym{g_{YM}}

\def\adot{\dot{a}}
\def\bdot{\dot{b}}
\def\bpa{\bar{\pa}}

\def\pr{\prime}
\def\ssk{\medskip}
\def\clb{\color{blue}}
\def\clr{\color{red}}
\def\clg{\color{green}}
\def\clp{\color{purple}}
\def\clc{\color{cyan}}
\def\clm{\color{magenta}}
\def\cly{\color{yellow}}

\def\bfA{{\bf A}}
\def\bfB{{\bf B}}
\def\bfK{{\bf K}}
\def\bfU{{\bf U}}
\def\bfX{{\bf X}}
\def\bfY{{\bf Y}}
\def\bfZ{{\bf Z}}
\def\bfg{{\bf g}}
\def\bfn{{\bf n}}

\def\bsk{\bigskip}
\def\ssk{\medskip}

\def\Ec{{\cal E}}

\begin{document}

\overfullrule=0pt
\parskip=2pt
\parindent=12pt
\headheight=0in \headsep=0in \topmargin=0in
\oddsidemargin=0in

\vspace{ -3cm}
\thispagestyle{empty}

 \vspace{0.1cm}

\setcounter{equation}{0}
\setcounter{footnote}{0}
\setcounter{section}{0}

\begin{center}

{\Large\bf  Finite-temperature renormalization of Standard Model coupled with gravity, and its implications for cosmology}

\vskip 0.8cm

%
%
I. Y. Park
\\
%
%
%
\vspace{0.3cm}
{\it Department of Applied Mathematics,
Philander Smith University 
                               \\
Little Rock, AR 72202, USA \\
inyongpark05@gmail.com
}

 \vspace{.5cm}

\end{center}

 \vspace{0.1cm}

\begin{abstract}

Finite-temperature one-loop renormalization of the Standard Model, coupled with dynamic metric, is conducted in this study. The entire analysis is coherently carried out by using the refined background field method, applied in the spirit of the Coleman-Weinberg technique. The general form of the propagator, introduced in our previous work to facilitate Feynman diagram computation in a general curved background, proves useful in the presence of time-dependent temperature. Its utilization allows for the renormalization analysis of a FLRW background to essentially reduce to that of a constant finite-T flat spacetime. For infrared physics, the actual curved background should be considered. The implications of our findings for cosmology, particularly the cosmological constant problem and Hubble tension, are discussed.

\end{abstract}
\newpage





\section{Introduction}

Quantum corrections, while often negligible in the study of astronomical and cosmological phenomena, manifest observable effects under certain circumstances. For such studies, one should, in principle, consider the quantum-level action, i.e., the 1PI action. Even when higher-derivative quantum corrections are deemed negligible, the renormalization of various constants, including the cosmological and Newton's constants, is generally necessary. Similarly, finite-temperature quantum field theory (QFT) \cite{Kapusta,Bellac, Laine:2016hma}  effects must be taken into account, as they, among other things, induce shifts in coupling constants, thus potentially leading to observable effects. However, until recently, the finite-T QFT effects have not been adequately addressed, at least not in the manner treated in the recent sequels of the present work, despite their a priori significant contributions to various physical quantities. In this work, we lay the foundations for a more systematic analysis of quantum-gravitational finite-temperature effects in the Standard Model \cite{Langacker}\cite{Schwartz} and the Standard Model of cosmology \cite{Weinberg,Gorbunov, Dodelson, Baumann}.

Essentially, we quantize the system of the Standard Model (SM) coupled to gravity by employing foliation-based quantization \cite{Park:2014tia,Park:2014qoa,Park:2015qxa,Park:2014noa,Park:2015ota,Park:2016zgt,Park:2018vci,Park:2019amz}, a method developed over recent years. It was shown that the physical states are three-dimensional and renormalizable. Here we conduct finite-temperature one-loop renormalization. Quantization of the SM in a flat background is well-established. The SM effective potential was computed to two loops long ago \cite{Ford:1992pn}. With the inclusion of gravity, the foliation-based approach\footnote{The foliation-based quantization scheme essentially established renormalizability of S-matrix of {\em the physical states} obtained by complete gauge-fixing. It does not contradict the well-known offshell non-renormalizability of gravity.} can be applied. This method has been successfully employed in analyzing a scalar-gravity system \cite{Park:2016zgt} as well as an Einstein-Maxwell system \cite{Park:2018vci}. In this work, we apply this method, in a streamlined and pedagogical manner, to the SM coupled with dynamic metric in finite temperature.

The presence of temperature introduces several complications. Notably, it breaks Lorentz symmetry, serving as an example of generalized spontaneous symmetry breaking: the vacuum no longer respects the symmetry. One uncharted complication introduced by the presence of temperature pertains to our ultimate goal of systematically analyzing finite-temperature effects in an FLRW background. The temperature of the FLRW universe is time-dependent, decreasing since the Big Bang. This raises the question of how to deal with finite-T effects since conventional finite-T QFT techniques, designed for dealing with constant temperatures, cannot be applied directly. As we will show, the key lies in the general form of the propagator introduced in \cite{Park:2021ohu} to expedite loop analysis in a curved background. With this device, one can essentially repeat the flat-spacetime constant-T QFT analysis.

UV divergences can be calculated by considering the zero-T case. The beta functions are primarily determined by $Z$ factors, which are not influenced by temperature in a renormalization scheme analogous to $\overline{MS}$.\footnote{In a more general renormalization scheme, the $Z$ factors will contain temperature-dependent terms. We will not consider such a scheme in the present work.} Therefore, the beta function results remain the same as in the zero-T case, allowing us to focus on the temperature-dependent finite terms. Once renormalization is completed, for certain cosmological studies such as the present one, one may consider the leading-order part of the 1PI action in the derivative expansion, which essentially corresponds to the renormalized action with quantum- and finite-T-corrected coupling constants.

In previous sequels \cite{Park:2021ohu}\cite{Park:2021vro}, finite-temperature contributions to the cosmological constant (CC) were analyzed in a scalar-gravity system. The renormalized mass of the scalar field was set to be on the order of the CMB temperature. Related ideas can be found in \cite{Balazs:2022anb} and \cite{Ageeva:2024qie}. The motivation behind this was to address the smallness of the CC, a well-known problem (see, e.g., \cite{Sola:2013gha} for a review of the CC problem). This step was further supported by the principle of minimal sensitivity \cite{Stevenson:1981vj}. In the early universe, during the period of high temperature, the CC is not expected to be small \cite{Weinberg} like its present value.\footnote{The CC will turn out to be big due to the finite-T contributions; see below.} This allows for the employment of an ordinary renormalization scheme akin to that of the zero-T analysis for the present work.

A delicate question arises regarding the application of the renormalization group. It concerns whether one can always consider, given one's chosen renormalization conditions, the entire range of the energy scale, from the CMB temperature scale to the Planck scale. It may not be possible, even in the absence of beyond-SM physics in the range. Instead, it may be necessary to consider certain renormalization conditions with a specific segment of the energy range. For instance, in addressing the low-energy and CC problem, it is acceptable to set the mass parameter to be small, on the order of the CMB temperature. Suppose one were to find the solutions to the beta functions. The question arises: would it be appropriate to apply these solutions for an energy scale such as the electroweak (EW) scale? The answer is likely to be negative. This is because the setup with the Higgs mass parameter set to be small would not be suitable, or at best, highly inconvenient for addressing physics at such a scale. At low energies, consideration of symmetry restoration is not necessary, hence setting the energy scale $\mu$ to be small does not pose any issues. However, this will not be the case for considering EW scale physics. In the present work, we consider a temperature substantially higher than the electroweak scale and adopt an $\overline{MS}$-like scheme, conceiving renormalization-group running from the electroweak energy scale to an energy scale several orders higher.

A notable technical aspect of our analysis is the position-space Feynman diagrammatic method called the refined background field method (RBFM) \cite{Park:2019amz}, applied in the spirit of the Coleman-Weinberg technique. (See, e.g., \cite{Sher:1988mj} for a review of the Coleman-Weinberg technique.) While for a one-loop potential of a relatively simple system, one can rely on the usual method of computing the determinant, this method isn't always applicable. Furthermore, what is often needed, as is the case here, is not only the potential itself but also the part of the effective action containing derivative terms.\footnote{The method is optimized for computing the 1PI action, unlike the usual momentum-space Feynman rules that are optimized for computing a scattering amplitude. One advantage of RBFM is that it allows for a mechanical procedure that can be applied to any theory, regardless of the availability of established momentum-space Feynman rules. In particular, the symmetry factor of a given diagram is automatically incorporated. We believe that the discrepancy found in section 2 between the literature and our present work is due to symmetry factors. Of course, one can compute the 1PI action using momentum-space rules if they are already established. However, even in that case, the procedure will be quite circuitous. (If one is interested in two-loop and higher-order loops, however, it would be more efficient to first establish the Feynman rules.)} Thus, we predominantly apply the RBFM. Additionally, the tool retains the power of the scalar background covariance even in the presence of temperature, making it a useful asset in handling IR divergences as well.

The results obtained in the main body have potentially interesting implications for the CC problem, the time dependence of effective coupling constants (especially, the cosmological and Newton's constants), and the Hubble tension \cite{Riess:2016jrr}\cite{Poulin:2018cxd}. While the crux argument for our CC-problem proposal was given in earlier sequels, we address a few loose ends here. In those sequels, the potential role of temperature in ameliorating the Hubble tension was also anticipated. (See \cite{Moshafi:2022mva} for a related discussion.) The finite-T effects must be observable in the CMB power spectrum analysis, as elaborated in the body.

\vspace{.2in}

The rest of the paper is organized as follows.

\vspace{.1in}

In section 2, we commence with a review of the zero-T flat spacetime one-loop renormalization of the SM. This section serves several purposes, primarily to establish the groundwork for the subsequent analyses, namely the examination of finite-T and graviton contributions. Notably, our computation of the Higgs coupling parameter, $\beta_\lambda$, diverges from the known literature; toward the conclusion of section 2, we address this discrepancy. In section 3, we delve into finite-T analysis, with a focus on the leading-temperature terms. After briefly addressing the UV divergences handled by the zero-T analysis, in section 3.1, we revisit the loop analysis, this time focusing on the IR divergences. Thermal resummation is carried out in both the gauge and fermion sectors in section 3.2, resulting in the acquisition of thermal masses for the gauge and fermion fields. In section 3.3 the Linde problem \cite{Linde:1980ts} (see \cite{Ghiglieri:2020dpq} for a review) is revisited, highlighting the potency of our formalism in addressing it. We employ a fermion-loop gauge 2-point amplitude to illustrate the problem and demonstrate that our temperature-driven renormalization scheme, incorporating thermal resummation in the gauge and fermion sectors, leads to non-vanishing magnetic thermal mass, the potential implication of which is discussed in the text. The renormalization scheme proposed in \cite{Park:2021ohu}, which we also adopt in this work, entails setting the Higgs mass $m_h$ to a value approximately around the EW scale, proving instrumental in managing the finite-T infrared divergence. In section 3.4, we undertake the task of enumerating the contributions of gravitons. While their contributions are anticipated to be negligible and not expected to alter the qualitative conclusions drawn from the SM sector analysis, consideration of a curved background is essential, given that a FLRW spacetime, which possesses a time-dependent temperature, {\em is} a curved background. Moving to section 4, we explore the implications of our findings for cosmology, particularly regarding Hubble tension and time-dependent effective coupling constants, following an elaboration on our previous CC resolution proposal. Concerning Hubble tension, the finite-T effect naturally hints at modifications to cosmological parameters, notably the Hubble constant. Numerical analysis, some preliminary results of which are presented, is underway to look deeper into this aspect. The ramifications of the present results for time-dependent effective coupling constants are intriguing, particularly the link between the temperature dependence of the effective CC and that of the effective Newton's constant, as implied by the quantum-level metric field equation. Finally, in section 5, we summarize our findings and outline several future directions for exploration.

\section{Zero-T flat spacetime renormalization}

In this section, we begin by examining the renormalization of the Standard Model (SM) in a zero-T flat background, laying the groundwork for subsequent, more complex finite-T analysis in a curved background. The renormalization process in a zero-T flat spacetime for the SM is well established, although there is some confusion, as far as we can tell, regarding the $\beta_\lambda$ result (see the remarks at the end of section 2.2). Ultraviolet divergences are determined by a zero-T flat background; however, introducing a nonzero temperature adds layers of complexity, especially in the context of the FLRW background later on. We utilize common steps for computing a given diagram in both zero-T and finite-T cases. Rather than solely presenting final results, we select several diagrams in the zero-T analysis and provide more detailed steps, which can then be applied in the subsequent finite-T analysis. With the zero-T divergences addressed in this section, our focus shifts to the temperature-dependent sectors in section 3. Throughout this process, we employ the refined background field method (RBFM), a review of which can be found in \cite{Park:2019amz}, and adopt a mostly plus metric.

\subsection{The setup}

The objective of both the present and subsequent sections is to derive the renormalized form of the action at finite temperature. This process unfolds in several steps. Initially, we revisit the analysis at zero temperature. The electroweak sector of the Standard Model action consists of the following components:
\be
{\cal L}={\cal L}_{gauge+gh}+{\cal L}_{fer}+{\cal L}_{Yukawa} +{\cal L}_{Higgs}.
\ee
Each sector is given as follows. The gauge sector is 
\bea
{\cal L}_{gauge}=-\fr14 W_{\m\n}^i W^{\m\n i}-\fr14 B_{\m\n} B^{\m\n}
\eea
with
\bea
W_{\m\n}^a = \pa_\m W_{\n}^a-\pa_\n W_{\m}^a+g f_{abc} W_{\m}^b W_{\n}^c\quad,\quad
B_{\m\n} = \pa_\m B_{\n}-\pa_\n B_{\m}
\eea
where $W_\m^a, B_\m$ denote $SU(2)\times U(1)$ gauge fields. The ghost action for the $SU(2)$ sector is
\bea
{\cal L}_{gh}=- \pa_\m c_a^* (\pa^\m c_a-g f_{abc}W_c^\m c_b).
\eea
The fermionic part of the Lagrangian reads:
\bea
{\cal L}_{fer} = -\sum_m^F\Big(  \barq_{mL}\, \slashed{D} q_{mL}+\barl_{mL}\, \slashed{D} l_{mL}
   +\baru_{mR}\,  \slashed{D} u_{mR}+\bard_{mR}\,  \slashed{D} d_{mR}+\bare_{mR}\,  \slashed{D} e_{mR}+\bar{\n}_{mR}\,  \slashed{D} \n_{mR}
\Big)
\eea
where
\bea
\bar{\psi}_{L,R}=(\psi_{L,R})^\dagger i\g^0\quad,\quad
\psi_{L,R} = P_{L,R}\,\psi \quad,\quad
P_L=\fr12 (1-\g^5)\quad,\quad P_R=\fr12 (1+\g^5).
\eea
The symbol $\psi$ collectively denotes the fermionic fields. The covariant derivatives of the fermions are given by
\bea
D_\m q_{mL\a}&=&\Big[\Big(\pa_\m I - \fr{i g}{2}\vec{\t}\cdot \vec{W}_\m -{ig'Y_L} B_\m\Big)\d_{\a\b}\Big]\left(
\begin{array}{c}
	u_{mL\b}  \\
	d_{mL\b}
\end{array}
\right) \nn\\
D_\m u_{mR\a}&=&\Big[\Big(\pa_\m I  -{ig'Y_R} B_\m\Big)\d_{\a\b}\Big]u_{mR\b}
\eea
where $\t$ denotes Pauli matrices. For the multiplet $(u_i,d_i,\n_i,e_i)^T$ where $i$ is the generation index, the values of the hypercharge are
\bea
Y_L = \Big(\fr16,\fr16,-\fr12,-\fr12\Big) \quad,\quad
Y_R = \Big(\fr23,-\fr13,0,-1\Big).
\eea
The quark Yukawa sector is given by
\bea
{\cal L}_{Yukawa}=-Y_{mn}^d\,\barq_L^{m}\Phi\, {d_R}^n
-Y_{mn}^u\,\barq_L^{m}\tilde{\F}\, u_R^n+h.c.
\eea
The covariant derivative of the scalar is given by
\bea
D_\m \F=\Big(\pa_\m -\fr{g}2 i\vec{\t}\cdot \vec{W}_\m -\fr{ig'}2  B_\m \Big)\F
\eea
with
\bea
\F=\left(
\begin{array}{c}
\F^+  \\
\F^0
\end{array}
\right)\quad,\quad
\tilde{\F}\equiv i\t_2 \F^\dagger=\left(
\begin{array}{c}
	\F^{0\dagger}  \\
	-\F^-
\end{array}
\right)\quad,\quad
\F^2\equiv \F^\dagger \F=\F^-\F^++\F^{0\dagger}\F^0  
\eea
where $\t_2=\left(
\begin{array}{cc}
0  & -i  \\
i  &  0
\end{array}
\right)$. The leptonic sector Lagrangian is given similarly. The Higgs sector is
\bea
{\cal L}_{Higgs} =  -  (D_\m \F)^\dagger (D^\m \F) -\fr{\l}{6}\Big(\F^2 {-} \fr{3}{\l}  {\tilde{\m}}^2\Big)^2  \la{cscalarL}
\eea
where the numerical coefficients of the potential term are chosen for canonical normalization of the kinetic and quartic potential terms of $\f$ that appear in the following parametrization,
\bea
\F=\left(
\begin{array}{c}
\upsilon^+  \\
\fr{\f+i\z}{\sqrt{2}}
\end{array}
\right) \la{Fdeff}
\eea
where $\upsilon^+$, $\upsilon^- (\equiv (\upsilon^+)^\dagger)$, $\z$ are Goldstone bosons. The above form of the parametrization will be explicitly used for some part of the analysis, such as Higgs mass analysis. With the parametrization, the Lagrangian \rf{cscalarL} reads
\bea
{\cal L}_{Higgs} = -\pa_\m \upsilon^- \pa^\m \upsilon^+
             -\fr12 \pa_\m \f\, \pa^\m \f -\fr12 \pa_\m \z\, \pa^\m \z
  -\fr{\l}{6}\Big[\fr12(2\upsilon^-\upsilon^++\f^2+\z^2) {-} \fr{3}{\l} \tilde{\m}^2\Big]^2
\la{grv-sclrq2}
\eea
For later purposes, let us compute the mass term by focusing on the $\f$ field. By shifting $\f \ra \f+\tilde{\m}\sqrt{\fr{6}{\l}}$ one gets, for the mass term,
\bea
 -\fr{\l}{6}\Big[\fr12\Big(\f+\tilde{\m}\sqrt{\fr{6}{\l}}\Big)^2  - \fr{3}{\l} \tilde{\m}^2\Big]^2
\ra  -\fr{\l}{6}\tilde{\m}^2{\fr{6}{\l}} \f^2= -\tilde{\m}^2 \f^2
=-\fr{m_h^2}2 \f^2 \la{masscomp}
\eea
where in the last equality we have defined
\be
m_h^2\equiv 2\tilde{\m}^2. \la{mbp}
\ee
After giving a vev to the Higgs field, the broken phase gauge fields are defined through
\bea
	Z_\m &\equiv& \cos\th_w W_\m^3- \sin\th_w B_\m \nn\\
	A_\m &\equiv& \sin\th_w W_\m^3+ \cos\th_w B_\m
\la{gaugerot}
\eea
where $\th_W$ denotes the Weinberg angle. One also introduces complex gauge fields:
\bea
W^+ = \fr{1}{\sqrt{2}}(W_\m^1-iW_\m^2)\quad,\quad
W^- = \fr{1}{\sqrt{2}}(W_\m^1+iW_\m^2).
\eea
As for the gauge group generators,
we adopt 
\bea
\sum_a \t_{ij}^a \t_{kl}^a=2 (\d_{il}\d_{kj}-\fr12 \d_{ij}\d_{kl})\quad,\quad
\tr(\t^a \t^b)=\tilde{C}_2 \d^{ab}\quad,\quad
\t^a \t^a= C_2 \bf{1}.
\eea
Above, $\tilde{C}_2=2$ and $C_2=3$ for $SU(2)$. We employ dimensional regularization, where an energy scale is introduced through the coupling constants in the following manner:
\bea
g\ra g \m^{\e},\quad g'\ra g' \m^{\e},\quad y_t\ra y_t \m^\e,\quad \l\ra \l \m^{2\e}
\eea
where $\e\equiv \fr{4-D}2$.
The coupling $y_t$ denotes top Yukawa coupling constant.
Let us illustrate the setup of RBFM with the Higgs scalar field. One first shifts
\be
\f\equiv \f_q+\f_B, 
\ee
where $\f_q,\f_B$ denote the quantum field and background field, respectively. (The subscript $q$ in $\f_q$ will be supressed in the body.) Conduct another shift
\be
\f_B\ra \f_B+\f_v \quad , \quad  \f_v\equiv \tilde{\m} \sqrt{\fr{6}{\l}}
\ee
where $\f_v$ denotes the vacuum expectation value, leading to
\be
\f\equiv \f_q+\vf\;\;\mbox{with}\;\; \vf\equiv \f_B+\f_v.  \la{rbfmdef}
\ee
As indicated, the field $\vf$ is a short-hand notation for $ \f_B+\f_v$; it can also be taken as an overall background field.
In other words one can conduct RBFM by treating $\vf$ as the background field, thus directly obtaining the 1PI action in terms of $\vf$ (and the gauge/fermion background fields). The renormalization constants, $Z$'s, are introduced according to
\be
W_0^\m\equiv \sqrt{Z_W}\,W^\m\quad,\quad
B_0^\m\equiv \sqrt{Z_B}\,B^\m\quad,\quad
\psi_0\equiv \sqrt{Z_\psi}\,\psi\quad,\quad
\Phi_0\equiv \sqrt{Z_\Phi}\,\Phi\quad,\quad
\ee
where the subscript $0$ denotes bare quantities. Similalry, for the coupling constants, one introduces
\be
g_0=Z_g g \quad,\quad
g_0'=Z_{g'} g' \quad,\quad
Y_0=Z_Y Y \quad,\quad
\l_0=Z_\l \l.
\ee
For each $Z$ we define $\D Z$ by
\be
Z\equiv 1+\D Z.
\ee

\vspace{.2in}

The computation of various diagrams below entails delicate technical subtleties, primarily due to the rotation of the gauge fields introduced in eq. 
\rf{gaugerot} and especially, the scalar's development of a vacuum expectation value. With the aim of computing the 1PI effective action, we initially employ the broken-phase action at a conceptual level. It is within this framework that the physical significance of various quantities, such as the renormalized Higgs mass parameter, can be properly identified. However, while one could, in principle, utilize the broken action for actual calculations, employing the original unbroken action is technically more straightforward and less tedious due to the preservation of $\vf$-covariance. For most computations, it is more efficient to utilize the original unbroken field basis. This consideration applies to all sectors, but it is particularly advantageous for the scalar sector. Below eq. \rf{vfsqres1}, we will illustrate this point with a scalar-loop two-point diagram.

\subsection{One-loop $\b$-functions}

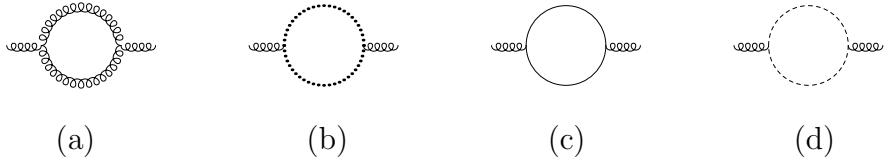
\begin{figure}
\centering
\begin{fmffile}{pure21}
	\!\!\Scale[0.35]{
		\begin{gathered}
		\begin{fmfgraph*}(160,100)
		\fmfleft{i} \fmfright{o}
		\fmf{gluon,tension=6}{i,v1} \fmf{gluon,tension=6}{v2,o}
		\fmf{gluon,left,tension=1.3}{v1,v2,v1}
		\end{fmfgraph*}
		\end{gathered}
	}
\end{fmffile}
\quad\quad
\begin{fmffile}{gauge2ptghL}
	\!\!\Scale[0.35]{
		\begin{gathered}
		\begin{fmfgraph*}(160,100)
		\fmfleft{i} \fmfright{o}
		\fmf{gluon,tension=6}{i,v1} \fmf{gluon,tension=6}{v2,o}
		\fmf{dots,width=3,left,tension=1.3}{v1,v2,v1}
		\end{fmfgraph*}
		\end{gathered}
	}
\end{fmffile}
\quad\quad
\begin{fmffile}{gauge2ptfL}
	\!\!\Scale[0.35]{
		\begin{gathered}
		\begin{fmfgraph*}(160,100)
		\fmfleft{i} \fmfright{o}
		\fmf{gluon,tension=6}{i,v1} \fmf{gluon,tension=6}{v2,o}
		\fmf{plain,left,tension=1.3}{v1,v2,v1}
		\end{fmfgraph*}
		\end{gathered}
	}
\end{fmffile}
\quad\quad
\begin{fmffile}{gauge2ptsL}
	\!\!\Scale[0.35]{
		\begin{gathered}
		\begin{fmfgraph*}(160,100)
		\fmfleft{i} \fmfright{o}
		\fmf{gluon,tension=6}{i,v1} \fmf{gluon,tension=6}{v2,o}
		\fmf{dashes,dash_len=4,left,tension=1.3}{v1,v2,v1}
		\end{fmfgraph*}
		\end{gathered}
	}
\end{fmffile}
\begin{center}
		\hspace{-.1in}	(a)\hspace{1.1in} \!\!(b) \hspace{.75in} \;\;\;\;(c) \hspace{.93in} \;(d)
	\end{center}
	\caption{Gauge wavefunction renormalization: (a) gauge loop (b) ghost loop (c) fermion loop (d) scalar loop}
	\label{gaugewfr}
\end{figure}

\begin{figure}
\centering
\begin{fmffile}{fermionrengf}
\!\!\Scale[0.45]{
\begin{gathered}
		\begin{fmfgraph*}(200,160)
			\fmfleft{i}
			\fmfright{o}
			\fmfbottom{b} 
			\fmf{plain,tension=1.5}{i,v1}
			\fmf{plain}{v1,v2}
			\fmf{plain,tension=1.5}{v2,o}
			\fmf{gluon,right,tension=1/5}{v2,v1}
			\fmffreeze
	\end{fmfgraph*}
	\end{gathered}
	}
\end{fmffile} 
\quad\quad\hspace{.5in}\;
\begin{fmffile}{fermionrensf}
\!\!\Scale[0.45]{
\begin{gathered}
		\begin{fmfgraph*}(200,160)
			\fmfleft{i}
			\fmfright{o}
			\fmfbottom{b} 
			\fmf{plain,tension=2}{i,v1}
			\fmf{plain}{v1,v2}
			\fmf{plain,tension=2}{v2,o}
			\fmf{dashes,right,tension=1/5}{v2,v1}
			\fmffreeze
	\end{fmfgraph*}
	\end{gathered}
	}
\end{fmffile} \\
\vspace{-.3in}
\!(a)\hspace{2in}\;\;\;(b)
\begin{center}
	\end{center}
	\vspace{-.4in}
	\caption{Fermion wavefunction renormalization}
	\label{fermionwfr}
\end{figure}
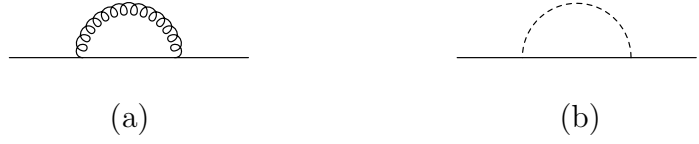

For the renormalization of the Higgs mass, it is necessary to first carry out the renormalization of the gauge and fermion fields. We commence with the wave-function renormalization of gauge fields. The relevant diagrams are given in Fig. \ref{gaugewfr}. To elucidate the computation of these diagrams, we provide pedagogical details for (a) and (b). Further elaboration on the refined background field method (RBFM) can be found in \cite{Park:2015ota,Park:2018vci,Park:2019amz}. The relevant vertices, following the background shift, stem from
\bea
-\fr14 W_{\m\n}W^{\m\n}
&\ra&-\fr{g^2}8  f_{abc} f_{a'b'c'}\Big\{
\w_{\m\n}^a W_b^\m W_c^\n\; \w_{\m'\n'}^{a'} W_{b'}^{\m'} W_{c'}^{\n'} \nn\\
&&\hspace{-1in}+4W_{\m\n}^a W_\m^b \w_\n^c\; W_{\m'\n'}^{a'} W_{\m'}^{b'} \w_{\n'}^{c'}
+4\w_{\m\n}^a W_b^\m W_c^\n\;W_{\m'\n'}^{a'} W_{\m'}^{b'} \w_{\n'}^{c'}
\Big\} \label{glg2pt}
\eea
where $\omega$ denote the background gauge field. Let us examine the contribution of the second term within the curly brackets for illustration purposes. After appropriate contractions of the quantum fields $W$'s, followed by Fourier transformation and evaluation of the loop momentum integration, the divergent part is obtained as follows:
\bea
-\fr{\G(\e)}{(4\pi)^2} g^2 f_{abc}f_{a'bc}\Big[\fr5{24} p_2^2\, \w^a(p_1)\cdot \w^{a'}(p_2)-\fr1{12} p_2\cdot \w^a(p_1)\,p_2\cdot \w^{a'}(p_2) \Big].
\eea
Above, $\G$ denotes gamma function and $\e\equiv \fr{4-D}{2}$; $p_1,p_2$ denote the momenta of the external background fields $\w$'s. This result cannot be expressed in a gauge-invariant form in position space. However, as anticipated, the sum of all three terms in eq. (\ref{glg2pt}), as well as the ghost diagram, can be shown to be gauge-invariant. By similarly computing the other two terms and the ghost loop diagram depicted in Fig. \ref{gaugewfr} (b), the sum is found to be
\bea
\begin{fmffile}{pure21}
	\!\!\Scale[0.2]{
		\begin{gathered}
		\begin{fmfgraph*}(160,100)
		\fmfleft{i} \fmfright{o}
		\fmf{gluon,tension=6}{i,v1} \fmf{gluon,tension=6}{v2,o}
		\fmf{gluon,left,tension=1.3}{v1,v2,v1}
		\end{fmfgraph*}
		\end{gathered}
	}
\end{fmffile}
+\;
\begin{fmffile}{gauge2ptghL3}
\!\!\Scale[0.3]{
		\begin{gathered}
		\begin{fmfgraph*}(100,60)
		\fmfleft{i} \fmfright{o}    
		\fmf{gluon,tension=6}{i,v1} 
		\fmf{gluon,tension=6}{v2,o}
		\fmf{dots,width=2,left,tension=1.3}{v1,v2,v1}
		\end{fmfgraph*}
	\end{gathered}
	}
\end{fmffile}
&=&\fr5{3}\fr{\G(\e)}{(4\pi)^2} g^2 \Big[ p_2^2\, \w^a(p_1)\cdot \w^{a}(p_2)- p_2\cdot \w^a(p_1)\,p_2\cdot \w^{a}(p_2) \Big]\nn\\
&=&\fr5{6}\fr{\G(\e)}{(4\pi)^2} g^2\int_x W_{\m\n}^a W^{\m\n a}
\eea
where $f_{abc}f_{a'bc}=2\d_{aa'}$ has been used, and in the second equality, $\omega$ is switched back to $W$ for notational convenience. The computation of the fermion diagram in Fig. \ref{gaugewfr} (c) will be crucial for illustrating our proposal toward resolving Linde problem in the finite-T context. For a fermionic loop, such as an SU(2) quark loop of a specific color and flavor, one obtains
\bea
\begin{fmffile}{gauge2ptfL}
	\!\!\Scale[0.18]{
		\begin{gathered}
		\begin{fmfgraph*}(160,100)
		\fmfleft{i} \fmfright{o}
		\fmf{gluon,tension=6}{i,v1} \fmf{gluon,tension=6}{v2,o}
		\fmf{plain,left,tension=1.3}{v1,v2,v1}
		\end{fmfgraph*}
		\end{gathered}
	}
\end{fmffile}&=& -g^2 \Ct_2 \int \fr{{\raisebox{0.05ex}{$\mathchar '26$}\mkern -9mu\delta}(p_l+p_2)}{k^2(k-p_2)^2}\Big[k^\m A^a_\m(p_1) \;(k-p_2)^\n A^a_\n(p_2)\nn\\
&&+(k-p_2)^\m A^a_\m(p_1) \;k^\n A^a_\n(p_2)
-k_\m (k-p_2)^\m \, A^{a\, \r}(p_1) A^a_\r(p_2)
\Big]  \la{g2ptfl}
\eea
where ${\raisebox{0.05ex}{$\mathchar '26$}\mkern -9mu\delta}(p_1+p_2)$ denotes $(2\pi)^4 \d(p_1+p_2)$. This step will be utilized in the subsequent finite-T computation. The total $SU(2)$ fermions, including quarks and leptons, yield
\bea
\begin{fmffile}{gauge2ptfL}
	\!\!\Scale[0.18]{
		\begin{gathered}
		\begin{fmfgraph*}(160,100)
		\fmfleft{i} \fmfright{o}
		\fmf{gluon,tension=6}{i,v1} \fmf{gluon,tension=6}{v2,o}
		\fmf{plain,left,tension=1.3}{v1,v2,v1}
		\end{fmfgraph*}
		\end{gathered}
	}
\end{fmffile}|_{\mbox{quarks + leptons}}
&=&- \fr{\tilde{C}_2}{2} \fr{\G(\e)}{(4\pi)^2} g^2
\int_x W_{\m\n}^a W^{\m\n a}.
\eea
The computation of the scalar loop diagram Fig. \ref{gaugewfr} (d) is summarized as follows:
\bea
\begin{fmffile}{gauge2ptsL2}
	\!\!\Scale[0.2]{
		\begin{gathered}
		\begin{fmfgraph*}(160,100)
		\fmfleft{i} \fmfright{o}
		\fmf{gluon,tension=6}{i,v1} \fmf{gluon,tension=6}{v2,o}
		\fmf{dashes,dash_len=4,left,tension=1.3}{v1,v2,v1}
		\end{fmfgraph*}
		\end{gathered}
	}
\end{fmffile}
&=& -\fr{\G(\e)}{(4\pi)^2} g^2 \fr{\tilde{C}_2}{48} \int_x W_{\m\n}^a W^{\m\n a}-\fr{\G(\e)}{(4\pi)^2} g^2 \fr{\tilde{C}_2}{4}  m_h^2 \int_x W_\m^a W^{a\m}
\la{wfrsl}
\eea
where $m_h$ denotes the mass of the broken phase, as given in \rf{mbp}. (In the present framework, based on the unbroken basis, the mass $m_h$ arises at the final stage when symmetry breaking is considered. See the comments below eq. \rf{potexpquad} and in the introduction of Section 3.) Note that a gauge field mass term is generated; refer to the remarks toward the end of this section. This mass term should be considered as part of the renormalization of the gauge boson mass. Noting $\tilde{C}_2=2$ and summing all of the contributions above, the total one-loop $W_{\mu\nu}^2$ term is
\bea
\fr{1}{48}\Big(40-48-2\Big)
\fr{\G(\e)}{(4\pi)^2} g^2  \int_x W_{\m\n}^a W^{\m\n a}
=  -\fr{5}{24}
\fr{\G(\e)}{(4\pi)^2} g^2 \int_x W_{\m\n}W^{\m\n}.
\eea
This implies
\be
\D Z_W=-\fr56 \fr{\G(\e)}{(4\pi)^2} g^2. \la{ZWvalue}
\ee
As for $Z_B$, by going through analogous steps one can show that the divergent part of the 1PI Lagrangian generated by various contributions is 
\bea
&&\Big(-\fr{2}{24}-\fr16-\fr53-\fr12-1\Big)\fr{\G(\e)}{(4\pi)^2} {g'}^2 \Big[ p_2^2 \,B_\m^a(p_1)B^{a\m}(p_2)-p_2 \cdot B^a(p_1)\,  p_2 \cdot B^a(p_2) \Big] \nn\\
&=&-\fr{41}{12} {g'}^2 \Big[ p_2^2 \,B_\m^a(p_1)B^{a\m}(p_2)-p_2 \cdot B^a(p_1)\,  p_2 \cdot B^a(p_2)
\Big]
\eea
where the front factor in the first line represents the contributions from the Higgs, doublet quarks, right-handed quarks, doublet leptons, right-handed leptons, respectively. From this it follows
\be
\Delta Z_B=-\fr{41}6 \fr{\G(\e)}{(4\pi)^2} g'^2. \la{deltaZB}
\ee
This completes the gauge wavefunction renormalization. Let us consider the fermion wavefunction renormalization. The relevant diagrams are in Fig. \ref{fermionwfr}.
Including the factor arising from the color charges one gets
\bea
\begin{fmffile}{fermionrengf}
\!\!\Scale[0.2]{
\begin{gathered}
		\begin{fmfgraph*}(200,160)
			\fmfleft{i}
			\fmfright{o}
			\fmfbottom{b} 
			\fmf{plain,tension=1.5}{i,v1}
			\fmf{plain}{v1,v2}
			\fmf{plain,tension=1.5}{v2,o}
			\fmf{gluon,right,tension=1/5}{v2,v1}
			\fmffreeze
	\end{fmfgraph*}
	\end{gathered}
	}
\end{fmffile}
=
\fr34\fr{\G(\e)}{(4\pi)^2} g^2\, \bar{q_L}  \g^\m \pa_\m q_L  \quad,\quad
\begin{fmffile}{fermionren2}
\!\!\Scale[0.4]{
\begin{gathered}
		\begin{fmfgraph*}(100,80)
			\fmfleft{i}
			\fmfright{o}
			\fmfbottom{b}
			\fmf{plain,tension=2}{i,v1}
			\fmf{plain}{v1,v2}
			\fmf{plain,tension=2}{v2,o}
			\fmf{dashes,right,tension=1/5}{v2,v1}
			\fmffreeze
	\end{fmfgraph*}
	\end{gathered}
	}
\end{fmffile}
= -\fr12 \fr{\G(\e)}{(4\pi)^2} y_t^2 \, \bar{q_L}  \g^\m \pa_\m q_L       \la{sctopsi}
\eea
where $C_2=3$ has been used; $y_t$ denotes the top Yukawa coupling.
From these one gets
\bea
\D Z_\psi=-\fr12 \fr{\G(\e)}{(4\pi)^2}\, y_t^2 +\fr34\fr{\G(\e)}{(4\pi)^2}  g^2.  \la{ferwavren}
\eea

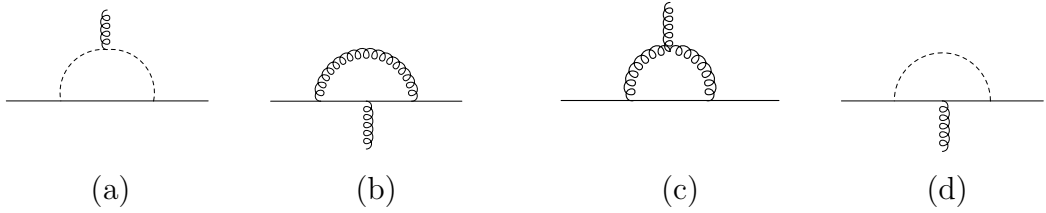
\begin{figure}
\centering
\begin{fmffile}{gaugeffssf}
\!\!\Scale[0.38]{
\begin{gathered}
\begin{fmfgraph*}(200,180) 
\fmfleft{i} \fmfright{o} 
\fmftop{t} 
\fmf{plain,tension=1.7}{i,v1} 
\fmf{plain}{v1,v2} 
\fmf{plain,tension=1.7}{v2,o} 
\fmffreeze 
\fmf{dashes,right=0.5,tension=1/2}{v2,v3,v1} \fmf{gluon,tension=1.3}{t,v3} 
\end{fmfgraph*} 
\end{gathered}
	}
\end{fmffile} 
\quad
\begin{fmffile}{collect3pt2}
\!\!\Scale[0.36]{
\begin{gathered}
		\begin{fmfgraph*}(200,100) 
\fmfleft{i} \fmfright{o} 
\fmfbottom{b} 
\fmf{plain,tension=1}{i,v1} 
\fmf{plain}{v1,v3,v2} 
\fmf{plain,tension=1}{v2,o} 
\fmf{gluon,right,tension=1/20}{v2,v1} 
\fmffreeze 
\fmf{gluon,tension=1/20}{b,v3} 
\end{fmfgraph*}  
	\end{gathered}
	}
\end{fmffile} 
\hspace{.3in}
\begin{fmffile}{collect3pt3}
\!\!\Scale[0.41]{
	\begin{gathered}
		\begin{fmfgraph*}(200,180) 
\fmfleft{i} \fmfright{o} 
\fmftop{t} 
\fmf{plain}{i,v1} 
\fmf{plain}{v1,v2} 
\fmf{plain}{v2,o} 
\fmffreeze 
\fmf{gluon,right=0.5,tension=1/2}{v2,v3,v1} \fmf{gluon,tension=1}{t,v3} 
\end{fmfgraph*} 
	\end{gathered}
	}
\end{fmffile}  
\quad
\begin{fmffile}{collect3pt4}
\!\!\Scale[0.38]{
\begin{gathered}
		\begin{fmfgraph*}(200,100) 
\fmfleft{i} \fmfright{o} 
\fmfbottom{b} 
\fmf{plain,tension=1}{i,v1} 
\fmf{plain}{v1,v3,v2} 
\fmf{plain,tension=1}{v2,o} 
\fmf{dashes,right,tension=1/20}{v2,v1} 
\fmffreeze 
\fmf{gluon,tension=1/20}{b,v3} 
\end{fmfgraph*}  
	\end{gathered}
	}
\end{fmffile} 
\begin{center}
\vspace{-.3in}
		\hspace{-.06in}	(a)\hspace{1.00in} \;\;\,(b) \hspace{1.15in} \;\;\,(c) \hspace{1.03in} \;(d)
	\end{center}
	\vspace{-.2in}
	\caption{Gauge coupling constant renormalization}
	\label{gaugeccr}
\end{figure}

As warm-up exercises (and also to verify the consistency of our method), let us compute the beta functions of the gauge sectors. The relevant diagrams are provided in Fig. \ref{gaugeccr}. (The fourth diagram above is relevant only for the case where the gauge line is that of $U(1)$.)
For $\beta_g$, one may consider either the $A\psi\psi$ vertex or the $AA\Phi\Phi$ vertex where $A$ ($\psi$) collectively denotes the gauge (fermionic) fields. The fact that they both yield the same result is guaranteed.\footnote{This is because if one appropriately rescales the gauge field by the (inverse) coupling constant, one obtains an action in which the coupling constant appears as an overall coefficient of the gauge kinetic terms. If one action is renormalizable, the other one should be too, since the two actions are related by a simple gauge field rescaling. The key point is that in the rescaled action, due to covariance, RBFM will produce terms involving $D_\mu \phi^\dagger D^\mu \phi$, so the coupling constant inside will remain unrenormalized, and the renormalization process will mainly affect the covariant kinetic terms. Therefore, in the rescaled action, $Z_A$ and $Z_g$ are not independently determined but only in a certain combination. Once one employs the original action, then $Z_A$ is determined. With this, $Z_g$ can be determined by utilizing the relation obtained in the rescaled action. (More specifically, consider the original action and determine the wavefunction renormalization. Then consider the coupling terms. Require that the $Z_g$ renormalization renders $D_\mu \phi^\dagger D^\mu \phi$ renormalized only by an overall factor of $Z_\phi$.) This, of course, must yield the same result as obtained by solely using the original action.} We choose the former for easier comparison with the literature. The result of the analysis is summarized as follows:
\bea
\begin{fmffile}{gaugeffssf}
\!\!\Scale[0.21]{
\begin{gathered}
\begin{fmfgraph*}(200,180) 
\fmfleft{i} \fmfright{o} 
\fmftop{t} 
\fmf{plain,tension=1.7}{i,v1} 
\fmf{plain}{v1,v2} 
\fmf{plain,tension=1.7}{v2,o} 
\fmffreeze 
\fmf{dashes,right=0.5,tension=1/2}{v2,v3,v1} \fmf{gluon,tension=1.3}{t,v3} 
\end{fmfgraph*} 
\end{gathered}
	}
\end{fmffile} 
&=& \fr14\fr{\G(\e)}{(4\pi)^2} igy_t^2 \, \barq_{L}\t^a \slashed{W}^a q_{L}
\nn\\
\begin{fmffile}{collect3pt2}
\!\!\Scale[0.20]{
\begin{gathered}
		\begin{fmfgraph*}(200,100) 
\fmfleft{i} \fmfright{o} 
\fmfbottom{b} 
\fmf{plain,tension=1}{i,v1} 
\fmf{plain}{v1,v3,v2} 
\fmf{plain,tension=1}{v2,o} 
\fmf{gluon,right,tension=1/20}{v2,v1} 
\fmffreeze 
\fmf{gluon,tension=1/20}{b,v3} 
\end{fmfgraph*}   
	\end{gathered}
	}
\end{fmffile}
&= &
-\fr18g^2 \fr{\G(\e)}{(4\pi)^2}   ig\,  \bar{q_L} {\t^a} \slashed{W}^a q_L
\nn\\
\begin{fmffile}{collect3pt3}
\!\!\Scale[0.22]{
	\begin{gathered}
		\begin{fmfgraph*}(200,180) 
\fmfleft{i} \fmfright{o} 
\fmftop{t} 
\fmf{plain}{i,v1} 
\fmf{plain}{v1,v2} 
\fmf{plain}{v2,o} 
\fmffreeze 
\fmf{gluon,right=0.5,tension=1/2}{v2,v3,v1} \fmf{gluon,tension=1}{t,v3} 
\end{fmfgraph*} 
	\end{gathered}
	}
\end{fmffile}  
&= &
\fr34g^2 \fr{\G(\e)}{(4\pi)^2}   ig\,  \bar{q_L} {\t^a} \slashed{W}^a q_L.
\eea
With these one gets
\bea
 \D Z_g
=-\fr{19}{12} \fr{\G(\e)}{(4\pi)^2} g^2. \la{deltaZg}
\eea
The beta function $\b_g$ is then given by
\bea
\b_g&=&\m \frac{d}{d \m}g= -\fr{1}{(4\pi)^2}  \fr{19}{6} g^3.
\eea
Let us turn to $\b_{g'}$. For an abelian gauge theory, it is well known that the coupling renormalization is determined by that of the wavefunction. The same applies to the abelian sector of the SM:
\bea
 \D Z_{g'} &=&-\fr12\D Z_B 
           =\fr{41}{12}\fr{\G(\e)}{(4\pi)^2} g'^2
\eea
where eq. \rf{deltaZB} has been used; one gets
\bea
\b_{g'}&=&\fr{1}{(4\pi)^2}  \fr{41}{6} g'^3.
\eea

\vspace{.2in}

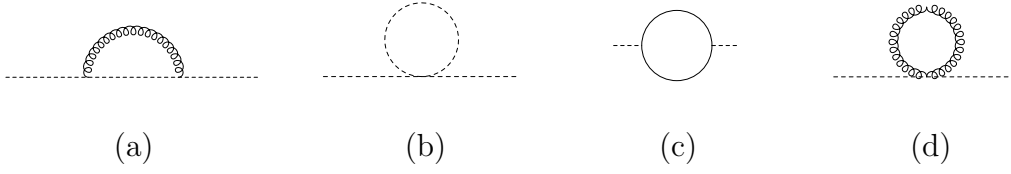
\begin{figure}
  \centering
\begin{fmffile}{gaugescalarsmalln}
\!\!\Scale[0.32]{
\begin{gathered}
		\begin{fmfgraph*}(300,250)
			\fmfleft{i}
			\fmfright{o}
			\fmfbottom{b}
			\fmf{dashes,tension=1.3}{i,v1}
			\fmf{dashes}{v1,v2}
			\fmf{dashes,tension=1.3}{v2,o}
			\fmf{gluon,right,tension=1/5}{v2,v1}
			\fmffreeze
	\end{fmfgraph*}
	\end{gathered}
	}
\end{fmffile} \quad
\begin{fmffile}{quartichiggssmall}
\!\!\Scale[0.37]{
	\begin{gathered}
		\begin{fmfgraph*}(200,150)
			\fmfleft{i}
			\fmfright{o}
			\fmftop{m}
			\fmf{dashes,tension=1}{i,v1}
			\fmf{dashes,tension=1}{v1,o}
			\fmf{dashes,left,tension=0}{v1,m,v1}
	       \end{fmfgraph*}
	\end{gathered}
	}
\end{fmffile}
\quad\quad
\begin{fmffile}{pure2small}
	\!\!\Scale[0.34]{
		\begin{fmfgraph*}(140,80)
		\fmfleft{i} \fmfright{o}
		\fmf{dashes,tension=5}{i,v1}
		\fmf{dashes,tension=5}{v2,o}
		\fmf{plain,left,tension=1}{v1,v2,v1}
		\end{fmfgraph*}
	}
\end{fmffile}
\quad\quad
\begin{fmffile}{s2ptgaugeloop}
\!\!\Scale[0.35]{
     \begin{gathered}
		\begin{fmfgraph*}(200,150)
			\fmfleft{i}
			\fmfright{o}
			\fmftop{m}
			\fmf{dashes,tension=1}{i,v1}
			\fmf{dashes,tension=1}{v1,o}
			\fmf{gluon,left,tension=0}{v1,m,v1}
	\end{fmfgraph*}
	\end{gathered}
	}
\end{fmffile}\\
\vspace{-.3in}
\hspace{0.1in} (a) \hspace{1.2in} (b) \hspace{1in} (c) \hspace{1.0in} (d)
 \caption{Higgs mass renormalization diagrams.}
\label{fig:mren}
\end{figure}
With the preparation above we now turn to the scalar mass renormalization with the diagrams given in Fig. \ref{fig:mren}. We quote the scalar sector Lagrangian,  eqs. \rf{cscalarL} and \rf{grv-sclrq2}, for convenience:
\bea
&& \hspace{.5in} {\cal L}_{Higgs} =  -  (D_\m \F)^\dagger (D^\m \F) -\fr{\l}{6}\Big(\F^2 {-} \fr{3}{\l} {\tilde{\m}}^2\Big)^2 \\
&=& -\pa_\m \upsilon^- \pa^\m \upsilon^+
             -\fr12 \pa_\m \f\, \pa^\m \f -\fr12 \pa_\m \z\, \pa^\m \z
  -\fr{\l}{6}\Big[\fr12(2\upsilon^-\upsilon^++\f^2+\z^2) {-} \fr{3}{{}\l} {\tilde{\m}}^2\Big]^2.
\la{grv-sclrq3} \nn
\eea
Let us calculate the diagram (b) by employing the real basis and Coleman-Weinberg technique. (Alternatively, one can, of course, apply the Feynman diagrammatic method that has been used so far.) For simplicity we demonstrate the computation by focusing on the background field $\vf$. The $\f$-quadratic part coming from the potential is
\bea
-\fr{\l}{6} \Big[ (\F^\dagger+\vF^\dagger)(\F+\vF)\Big]^2
\Rightarrow-\fr{\l}{6} \Big[ \fr{(\f+\vf)^2+2\upsilon^+\upsilon^-+\z^2 }{2}\Big]^2
\Rightarrow
 \fr{\l}{6} \Big[ \fr32\vf^2\f^2+\vf^2 \upsilon^+\upsilon^- +\fr12\vf^2\z^2 \Big].\nn\\
 \la{potexpquad}
\eea
Here, $\vf$ represents the background field, defined as $\vf \equiv \phi_B + \phi_v$ as per eq. \rf{rbfmdef}; $\phi$, $\upsilon^\pm$, and $\zeta$ denote the quantum fields. The point previously mentioned regarding the broken- vs. unbroken-phase calculation can be illustrated here. Although the analyses for the broken and unbroken phases are closely related, it is more efficient to compute the 1PI action for $\vf$ (using the unbroken-phase fields for the external fields) even after the shift $\phi \rightarrow \phi + \phi_v$. Applying the Coleman-Weinberg technique, one obtains, for its contribution to the 1PI action:
\bea
\ra \fr{\G(\e)}{(4\pi)^2} \Big(1+\fr23+\fr13\Big) \fr{\l}{4} m_h^2\, \vf^2
=\fr{\G(\e)}{(4\pi)^2}\, \fr{\l}{2}m_h^2\;\vf^2  \la{vfsqres1}
\eea
where, $(1, \frac{2}{3}, \frac{1}{3})$ represent the contributions from $(\phi, \upsilon, \zeta)$, respectively. This constitutes the $\phi$-part of the complete $\lambda m_h^2 \, \Phi^\dagger \Phi$. (To derive the full expression, $\lambda m_h^2\, \Phi^\dagger \Phi$, one must also consider the background scalar fields of $\upsilon$ and $\zeta$.) Let's verify the result by evaluating the diagram in the complex basis. The quartic Higgs coupling diagram (b) yields, in the complex field basis:
\bea
- \l\fr{(\F^\dagger \F)^2}{6}
\Rightarrow -\l\int d^4x \, \vF^\dagger \vF \int   \fr{d^4k}{(2\pi)^4}\fr{1}{k^2+m_h^2}.
\eea
Evaluating the Euclidean momentum loop integration, one gets
\bea
&\Rightarrow& -\l\int d^4x \, \vF^\dagger \vF \int   \fr{d^4k}{(2\pi)^4}\fr{1}{k^2+m_h^2}
=i \l\fr{\G(\e)}{(4\pi)^2}m_h^2 \int_x \vF^\dagger \vF.
\la{zeroT2ptsl}
\eea
The contribution to the 1PI action $\G$ is then given by
\bea
\begin{fmffile}{quartichiggssmall}
\!\!\Scale[0.25]{
	\begin{gathered}
		\begin{fmfgraph*}(200,150)
			\fmfleft{i}
			\fmfright{o}
			\fmftop{m}
			\fmf{dashes,tension=1}{i,v1}
			\fmf{dashes,tension=1}{v1,o}
			\fmf{dashes,left,tension=0}{v1,m,v1}
	       \end{fmfgraph*}
	\end{gathered}
	}
\end{fmffile}
= \fr{\G(\e)}{(4\pi)^2} \l m_h^2 \int_x \vF^\dagger \vF
 \la{hmcbp}
\eea
which, for the $\vf$-dependent part, is the same as \rf{vfsqres1}. The remaining diagrams can be summarized as
\bea
\begin{fmffile}{gaugescalarsmall}
\!\!\Scale[0.12]{
\begin{gathered}
		\begin{fmfgraph*}(300,250)
			\fmfleft{i}
			\fmfright{o}
			\fmfbottom{b} 
			\fmf{dashes,tension=1.3}{i,v1}
			\fmf{dashes}{v1,v2}
			\fmf{dashes,tension=1.3}{v2,o}
			\fmf{gluon,right,tension=-.5}{v2,v1}
			\fmffreeze
	\end{fmfgraph*}
	\end{gathered}
	}
\end{fmffile}
&=&
\fr{\G(\e)}{(4\pi)^2}(g^2 C_2+g'^2)\fr{1}{2}\int_x\pa^\m \vF^\dagger  \pa_\m \vF
-\fr{\G(\e)}{(4\pi)^2}(g^2 C_2+g'^2)\fr{m_h^2}{4}\int_x  \vF^\dagger \vF  \nn\\
\begin{fmffile}{pure2small}
	\!\!\Scale[0.25]{
	\begin{gathered}
		\begin{fmfgraph*}(140,80)
		\fmfleft{i} \fmfright{o}
		\fmf{dashes,tension=5}{i,v1} 
		\fmf{dashes,tension=5}{v2,o}
		\fmf{plain,left,tension=1}{v1,v2,v1}
		\end{fmfgraph*}
  \end{gathered}
	}
\end{fmffile}  
&=&-3 y_t^2  \fr{\G(\e)}{(4\pi)^2}\int_x  \pa^\m \vF^\dagger  \pa_\m \vF \quad,\quad
\begin{fmffile}{s2ptgaugeloop}
\Scale[0.26]{
     \begin{gathered}
		\begin{fmfgraph*}(200,150)
			\fmfleft{i}
			\fmfright{o}
			\fmftop{m}
			\fmf{dashes,tension=1}{i,v1}
			\fmf{dashes,tension=1}{v1,o}
			\fmf{gluon,left,tension=0}{v1,m,v1}
	\end{fmfgraph*}
	\end{gathered}
	}
\end{fmffile}  =0.  \la{higgs2ptcoll}
\eea
The gauge-loop diagram above is considered to vanish. This demonstrates the efficacy of applying the RBFM in the spirit of Coleman-Weinberg. These gauge mass-dependent terms can be disregarded because we are considering the $Z$'s for the unbroken action. At zero temperature, we can primarily rely on the unbroken basis, and that suffices.\footnote{In other words, it is unnecessary to explicitly consider the Higgs shift to ensure the removal of UV divergences for all non-Higgs sectors. For the Higgs sector on the other hand, it is crucial to focus on the proper Higgs mass term and consider the shift, since our interest lies not in the renormalization of the wrong-sign mass term but in that of the proper Higgs mass term of the broken phase. Consideration of the proper mass term amounts to employing the massive Higgs propagator.} The wave-function renormalization constant $\Delta Z_\Phi$ is determined by
\bea
 -\D Z_\F\pa_\m\F^\dagger \pa^\m \F  +\fr{\G(\e)}{(4\pi)^2}(g^2 C_2+g'^2)\fr{1}{2}\int \pa^\m \F^\dagger \pa_\m \F 
  -3 y_t^2 \fr{\G(\e)}{(4\pi)^2}\int \pa^\m \F^\dagger\pa_\m \F  =0
\eea
which yields
\be
\D Z_\F=\fr{\G(\e)}{(4\pi)^2}\Big[\fr{1}{2}(g^2 C_2+g'^2)  -3 y_t^2\Big].
\ee
The mass renormalization goes:
\bea
-m_h^2 (\D Z_{m_h^2}+\D Z_{\F})  \vF^\dagger \vF  -\fr{\G(\e)}{(4\pi)^2}(g^2 C_2+g'^2)\fr{m_h^2}{4}\vF^\dagger\vF
+m_h^2{\l} \fr{\G(\e)}{(4\pi)^2}      \vF^\dagger  \vF
=0,
\la{massshifteq}
\eea
leading to
\bea
\D Z_{m_h^2}=\fr{\G(\e)}{(4\pi)^2}\Big[-\fr34(g^2 C_2+g'^2)+ \l  +3 {y_t^2} \Big].
\eea
The beta function then reads
\bea
\b(m_h^2)
       = \fr{m_h^2}{(4\pi)^2} \Big[-\fr9{2} g^2 -\fr3{2} g'^2+{2} \l  +{6} y_t^2 \Big].  \la{hmbf}
\eea
This result confirms the corresponding result of Ford et al. (1992) \cite{Ford:1992pn}.

\vspace{.2in}

\begin{figure}
\centering

\begin{fmffile}{higgs4ptgl}
\!\!\Scale[0.3]{
\begin{fmfgraph*}(200,140)
\fmfleft{i1,i2}
\fmfright{o1,o2}
\fmfbottom{bi,b1,b2,b3,b4,bf}
\fmf{dashes,tension=2}{i1,v1,i2}
\fmf{dashes,tension=2}{o1,v2,o2}
\fmf{gluon,left=1}{v1,v2,v1}
\fmffreeze
\end{fmfgraph*}
                }
\end{fmffile}
\quad
\begin{fmffile}{higgs4pt1}
\!\!\Scale[0.3]{
		\begin{fmfgraph*}(180,140)
			\fmfleft{i1,i2}
\fmfright{o1,o2}
\fmf{dashes}{i1,v1}
\fmf{plain,tension=.5}{v1,v3}
\fmf{dashes}{v3,o1}
\fmf{dashes}{o2,v4}
\fmf{plain,tension=.5}{v4,v2}
\fmf{dashes}{v2,i2}
\fmf{plain,tension=.5}{v1,v2}
\fmf{plain,tension=.5}{v3,v4}
	\end{fmfgraph*}
	              }
\end{fmffile}
\begin{fmffile}{higgs4ptggss}
\!\!\Scale[0.3]{
		\begin{fmfgraph*}(200,160)
			\fmfleft{i1,i2}
\fmfright{o1,o2}
\fmf{dashes}{i1,v1}
\fmf{gluon,tension=.5}{v1,v3}
\fmf{dashes}{v3,o1}
\fmf{dashes}{o2,v4}
\fmf{gluon,tension=.5}{v4,v2}
\fmf{dashes}{v2,i2}
\fmf{dashes,tension=.7}{v1,v2}
\fmf{dashes,tension=.7}{v3,v4}
	\end{fmfgraph*}
	              }
\end{fmffile} 
\quad
\\
\hspace{-.15in}	(a)\hspace{.9in} \!\!(b) \hspace{.6in}\; (c)\\
\vspace{.1in}
\begin{fmffile}{higgs4ptggs} 
\!\!\Scale[0.3]{
\begin{fmfgraph*}(200,140) 
\fmfleft{i1,i2} 
\fmfright{o1,o2} 
\fmf{dashes,tension=2.2}{i1,g1} 
\fmf{dashes,tension=2.2}{g2,o1} 
\fmf{dashes,tension=3.2}{i2,v1,o2} 
\fmf{gluon,tension=1.2}{v1,g1} 
\fmf{gluon,tension=1.2}{v1,g2} 
\fmf{dashes,tension=0.5}{g1,g2} 
\end{fmfgraph*} 
               }
\end{fmffile} 
\;\;
\begin{fmffile}{higgs4ptgss} 
\!\!\Scale[0.3]{
\begin{fmfgraph*}(180,120) 
\fmfleft{i1,i2} 
\fmfright{o1,o2} 
\fmf{dashes,tension=2.2}{i1,g1} 
\fmf{dashes,tension=2.2}{g2,o1} 
\fmf{dashes,tension=3.2}{i2,v1,o2} 
\fmf{dashes,tension=1.2}{v1,g1} 
\fmf{dashes,tension=1.2}{v1,g2} 
\fmf{gluon,tension=0.5}{g1,g2} 
\end{fmfgraph*} 
               }
\end{fmffile}
\begin{fmffile}{higgs4ptsl} 
\!\!\Scale[0.3]{
\begin{fmfgraph*}(200,140) 
\fmfleft{i1,i2} 
\fmfright{o1,o2} 
\fmfbottom{bi,b1,b2,b3,b4,bf} 
\fmf{dashes,tension=2}{i1,v1,i2} 
\fmf{dashes,tension=2}{o1,v2,o2} 
\fmf{dashes,left=1}{v1,v2,v1} 
\fmffreeze 
\end{fmfgraph*} 
                }
\end{fmffile}
\begin{center}
\hspace{0in}\!\! (d) \hspace{.8in} (e) \hspace{.7in}\! (f)
	\end{center}
	\caption{Higgs coupling renormalization. The diagrams with external lines permuted are not included.}
	\vspace{.3in}
	\label{lambdaren}
\end{figure}
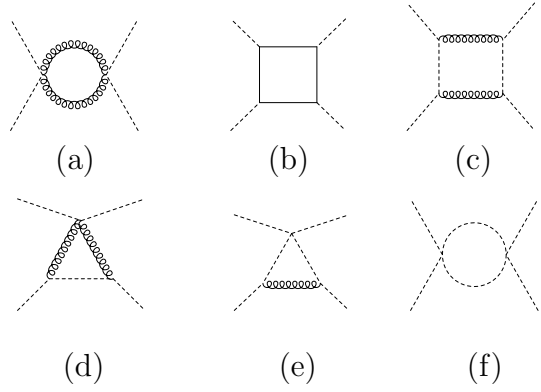

For the renormalization of the Higgs quartic coupling, the relevant graphs are provided in Fig. \ref{lambdaren}. (The diagrams obtained by permuting the external lines - which are not shown - are associated with the symmetry factors, which are automatically accounted for in the RBFM formalism.)
We illustrate the calculation with diagrams (a) and (b). For diagram (a), there are three different kinds, depending on which of the $SU(2)\times U(1)$ gauge fields runs on the loop. Let us consider the case of $W$-fields running on the loop. After simplification of the initial correlator expression, we obtain
\bea
\begin{fmffile}{higgs4ptgl2}
\hspace{-.05in}\Scale[0.4]{
       \begin{gathered}
		\begin{fmfgraph*}(70,50)
\fmfleft{i1,i2}
\fmfright{o1,o2}
\fmfbottom{bi,b1,b2,b3,b4,bf}
\fmf{dashes,tension=2}{i1,v1,i2}
\fmf{dashes,tension=2}{o1,v2,o2}
\fmf{gluon,left=1}{v1,v2,v1}
\fmffreeze
	\end{fmfgraph*}
	  \end{gathered}
	              }
\end{fmffile}
&=&\fr34 g^4\int_{p's,k}
\Big[\vF(p_1)^\dagger\vF(p_2)\Big]\Big[\vF^\dagger(p_3) \vF(p_4)\Big]\;
\fr{{\raisebox{0.05ex}{$\mathchar '26$}\mkern -9mu \delta}(\Sigma p_l)}{k^2 (k+p_3-p_4)^2 }
\la{h4ptgloop}
\eea
where ${\raisebox{0.05ex}{$\mathchar '26$}\mkern -9mu \delta}(\Sigma p_l)$ denotes $(2\pi)^4 \d(p_1+p_3-p_2-p_4)$. The finite-T counterpart of this expression will serve as the starting point for the same diagram in the finite-T analysis in section 3. Performing the momentum integrations yields
\bea
\begin{fmffile}{higgs4ptgl2}
\hspace{-.05in}\Scale[0.4]{
       \begin{gathered}
		\begin{fmfgraph*}(70,50)
\fmfleft{i1,i2}
\fmfright{o1,o2}
\fmfbottom{bi,b1,b2,b3,b4,bf}
\fmf{dashes,tension=2}{i1,v1,i2}
\fmf{dashes,tension=2}{o1,v2,o2}
\fmf{gluon,left=1}{v1,v2,v1}
\fmffreeze
	\end{fmfgraph*}
	  \end{gathered}
	              }
\end{fmffile}
=
\fr{\G(\e)}{(4\pi)^2}\fr34 g^4 \int_x (\vF^\dagger \vF)^2.
\eea
Similarly, for diagram (b), one can show, after some algebra,
\bea
\begin{fmffile}{higgs4pt13}
\Scale[0.4]{
       \begin{gathered}
		\begin{fmfgraph*}(70,50)
			\fmfleft{i1,i2}
\fmfright{o1,o2}
\fmf{dashes}{i1,v1}
\fmf{plain,tension=.5}{v1,v3}
\fmf{dashes}{v3,o1}
\fmf{dashes}{o2,v4}
\fmf{plain,tension=.5}{v4,v2}
\fmf{dashes}{v2,i2}
\fmf{plain,tension=.5}{v1,v2}
\fmf{plain,tension=.5}{v3,v4}
	\end{fmfgraph*}
	     \end{gathered}
	              }
\end{fmffile}
&=&-{y_t^4}(\d^{\m\n}\d^{\r\s}-\d^{\m\r}\d^{\n\s}+\d^{\m\s}\d^{\n\r})\int_{p's,k}
\vF_m(p_1)\vF_n(p_2)\vF_m^\dagger(p_3) \vF_n^\dagger(p_4)\; {\raisebox{0.05ex}{$\mathchar '26$}\mkern -9mu\delta}(\Sigma p_l)\nn\\
&&\qquad\fr{k_\m k_\n k_\r k_\s}{k^2 (k-p_1)^2(k-p_1+p_4)^2 (k-p_3)^2}  \la{h4ptfloop}
\eea
where ${\raisebox{0.05ex}{$\mathchar '26$}\mkern -9mu\delta}(\Sigma p_l)$ denotes $(2\pi)^4 \d(p_1+p_2-p_3-p_4)$. As in the previous diagram, an analogous expression will be the starting point of the same diagram in the finite-temperature analysis. Performing the momentum integrations one gets
\bea
\begin{fmffile}{higgs4pt13}
\Scale[0.4]{
       \begin{gathered}
		\begin{fmfgraph*}(70,50)
			\fmfleft{i1,i2}
\fmfright{o1,o2}
\fmf{dashes}{i1,v1}
\fmf{plain,tension=.5}{v1,v3}
\fmf{dashes}{v3,o1}
\fmf{dashes}{o2,v4}
\fmf{plain,tension=.5}{v4,v2}
\fmf{dashes}{v2,i2}
\fmf{plain,tension=.5}{v1,v2}
\fmf{plain,tension=.5}{v3,v4}
	\end{fmfgraph*}
	     \end{gathered}
	              }
\end{fmffile}
&=& - \fr{\G(\e)}{(4\pi)^2}\, 3y_t^4 \, \int_x (\vF^\dagger \vF)^2
\eea
where the factor 3 comes from the color index of the fermions. Diagrams (c) - (e) can be similarly computed:
\bea
&&
\hspace{-.4in}
\begin{fmffile}{higgs4ptggss2}
\hspace{-.05in}\Scale[0.4]{
       \begin{gathered}  		
		\begin{fmfgraph*}(70,50)	
\fmfleft{i1,i2}
\fmfright{o1,o2}
\fmf{dashes}{i1,v1}
\fmf{gluon,tension=.5}{v1,v3}
\fmf{dashes}{v3,o1}
\fmf{dashes}{o2,v4}
\fmf{gluon,tension=.5}{v4,v2}
\fmf{dashes}{v2,i2}
\fmf{dashes,tension=.7}{v1,v2}
\fmf{dashes,tension=.7}{v3,v4}
	\end{fmfgraph*}
	  \end{gathered}  
	              }
\end{fmffile}
=\fr{\G(\e)}{(4\pi)^2}\fr1{16} g^4 \int_x (\vF^\dagger \vF)^2,
\qquad 
\begin{fmffile}{higgs4ptggs2}
\hspace{-.05in}\Scale[0.4]{
       \begin{gathered}
		\begin{fmfgraph*}(70,50)
\fmfleft{i1,i2}
\fmfright{o1,o2}
\fmf{dashes,tension=2.2}{i1,g1}
\fmf{dashes,tension=2.2}{g2,o1}
\fmf{dashes,tension=3.2}{i2,v1,o2}
\fmf{gluon,tension=1.2}{v1,g1}
\fmf{gluon,tension=1.2}{v1,g2}
\fmf{dashes,tension=0.5}{g1,g2}
	\end{fmfgraph*}
	  \end{gathered}
	              }
\end{fmffile}
=
-\fr{\G(\e)}{(4\pi)^2}\fr58 g^4 \int_x (\vF^\dagger \vF)^2 \nn\\
&&
\hspace{.5in}
\begin{fmffile}{higgs4ptgss2}
\hspace{-.05in}\Scale[0.4]{
       \begin{gathered}  		
		\begin{fmfgraph*}(70,50)	
\fmfleft{i1,i2} 
\fmfright{o1,o2} 
\fmf{dashes,tension=2.2}{i1,g1} 
\fmf{dashes,tension=2.2}{g2,o1} 
\fmf{dashes,tension=3.2}{i2,v1,o2} 
\fmf{dashes,tension=1.2}{v1,g1} 
\fmf{dashes,tension=1.2}{v1,g2} 
\fmf{gluon,tension=0.5}{g1,g2} 
	\end{fmfgraph*}
	  \end{gathered}  
	              }
\end{fmffile}
= -\fr{\G(\e)}{(4\pi)^2}\fr{\l_s}{4 } g^2 \int_x (\vF^\dagger\vF)^2. 
\eea
Finally, consider Fig. \ref{lambdaren} (f). By employing the complex basis Feynman diagram method one can show
\bea
\begin{fmffile}{higgs4ptsl2}
\hspace{-.05in}\Scale[0.4]{
       \begin{gathered}  		
		\begin{fmfgraph*}(70,50)	
\fmfleft{i1,i2} 
\fmfright{o1,o2} 
\fmfbottom{bi,b1,b2,b3,b4,bf} 
\fmf{dashes,tension=2}{i1,v1,i2} 
\fmf{dashes,tension=2}{o1,v2,o2} 
\fmf{dashes,left=1}{v1,v2,v1} 
\fmffreeze 
	\end{fmfgraph*}
	  \end{gathered}  
	              }
\end{fmffile}
 =\fr{\G(\e)}{(4\pi)^2}\fr{\l^2}{3} \int_x (\varPhi^\dagger \varPhi)^2.  
 \la{lambdadiv}
\eea
Let us cross-check this result, this time, by employing the real basis Coleman-Weinberg technique. As before, we focus on the $\vf$ background field; the relevant $\f$-quadratic part of the potential is given in \rf{potexpquad}. Each of these vertices on the right-hand side yields $\vf^4$ term with the following sum:
\bea
\fr{\l^2}{12}\fr{\G(\e)}{(4\pi)^2}\int_x \vf^4
\eea
which confirms \rf{lambdadiv} for the $\vf^4$ part.

\vspace{.2in}
One is now ready to tackle renormalization of $\l$. Adding all of the contributions from Fig. \ref{lambdaren} (a)-(f) (and the conributions from the $U(1)$ gauge field as well), one gets 
\bea
\hspace{-.2in}\fr{(4\pi)^2}{\G(\e)}{4}\l \D Z_{\l}
 &=& 8\l^2 -18\l g^2 -6\l g'^2  +24 \l y_t^2 + \fr9{2}g^4+3g^2g'^2+\fr{3}{2}g'^4-72 y_t^4.
\eea
Requiring the bare coupling's independence of the scale,
\bea
\hspace{-1in}&&\m \fr{d}{d\m} (\m^{ 2\e}Z_{\l} \l)=0
\quad\Rightarrow \quad
\fr{\m}{\l}  \fr{d}{d\m}\l
       =-\fr{\m}{Z_{\l}}\fr{d}{d\m}Z_{\l}\simeq -\m\fr{d}{d\m}Z_{\l}   \la{higgsbareren}
\eea
where the last equality is valid in the one-loop approximation. From this it follows
\bea
\hspace{-.2in} \m\fr{d}{d\m}Z_{\l}
&=&\fr{\G(\e)}{(4\pi)^2} \m\fr{d}{d\m} \Big[2\l -\fr92 g^2 -\fr32 g'^2  +6  y_t^2 + \fr9{8}\fr{g^4}{\l}+\fr34\fr{g^2g'^2}{\l}+\fr{3}{8}\fr{g'^4}{\l}-18 \fr{y_t^4}{\l} \Big].
\eea
The Higgs coupling beta function is then given by
\bea
\b_\l&\equiv& \m\frac{d}{d \m}\l = -\l \m\frac{d}{d \m} Z_\l\nn\\
&\simeq& \fr{1}{(4\pi)^2} \Big[{4}\l^2 -9 \l g^2 -3\l g'^2   +12  \l y_t^2         +\fr9{{4}} {g^4} +\fr3{{2}} {g^2g'^2} +\fr3{4}{g'^4}-36 {y_t^4}\Big]. \label{mybetaL}
\eea
Some of the coefficients above do not agree with those of Machacek and Vaughn (1984) \cite{Machacek:1984zw}, as noted in the remarks below.

\vspace{.2in}

\ni {\bf Remarks:} \\ \vspace{-.2in}

\ni 
Let us revisit the following aspects of the analysis above. We have observed that the gauge two-point function with a Higgs loop generates a gauge mass term. In QED, the photon self-energy term does not violate gauge invariance; namely, the photon mass term is not generated. As we have noted, this is not the case for the SM. This may not be a problem for the SM, at least not a serious one, since the gauge mass term comes from the Higgs mass term, which in turn originates from the spontaneous symmetry breaking: the gauge symmetry is broken anyway. However, this could pose a problem for a scalar QED: although the classical action is gauge invariant, the one-loop self-energy is not. The takeaway should be that the usefulness of a regularization method is theory-dependent.

As for the result \rf{mybetaL}, let us compare it with the corresponding result in \cite{Machacek:1984zw}\cite{Pirogov:1998tj} where a different convention, $-\fr{\l}{2}\Big(\F^2 {-} \fr{3}{\l} { \tilde{\m}}^2\Big)^2$, was employed for the quartic potential term:
\bea
(4\pi)^2\b_\l &=& 12 \l^2-(3 {g'}^2+9g^2)\l+\fr94 g^4+\fr32 g^2g'^2+\fr34 g'^4+\cdots. \la{machacekresult}
\eea
To make comparison, let us go back to the right-hand side of \rf{higgsbareren} and rescale it:
\bea
\Rightarrow \qquad
\fr{\m}{\l}  \fr{d}{d\m}\l = -\m\fr{d}{d\m}Z_{3\l}
\eea
from which one gets
\bea
&&\hspace{1.1in} \b_\l=\m  \fr{d}{d\m}\l = -\m \l\fr{d}{d\m}Z_{3\l}\nn\\
&&\hspace{-.5in}=\;\fr{\G(\e)}{(4\pi)^2} \m\fr{d}{d\m} \Big[6\l -\fr92 g^2 -\fr32 g'^2  +6  y_t^2 + \fr3{8}\fr{g^4}{\l}+\fr14\fr{g^2g'^2}{\l}+\fr{1}{8}\fr{g'^4}{\l}-6 \fr{y_t^4}{\l} \Big].
\eea
This leads to
\bea
(4\pi)^2\b_\l &=& 12 \l^2-(3 {g'}^2+9g^2)\l+\fr34 g^4+\fr12 g^2g'^2+\fr14 g'^4+\cdots.  \la{vaughnresult}
\eea
This does not agree with the existing result, eq. \rf{machacekresult}. For instance, the coefficients of the quartic gauge coupling terms in \rf{machacekresult} are larger than those in (\ref{vaughnresult}) by a factor 3.

\section{Finite-temperature renormalization in a curved spacetime}

With the completion of the zero-T flat spacetime analysis, in this section, we branch out to investigating the constant finite-T effects of the Standard Model (SM) in both a flat and FLRW background. To accurately account for temperature effects in cosmology, it is essential to consider not only the thermodynamics of the hydrodynamic matter but also the finite-T quantum-field-theoretic contributions of the SM fields. We analyze the finite-temperature shifts of coupling constants, combination of which with the original coupling constants should be regarded as effective coupling "constants." This entrance of effective coupling constants is exemplified by the Higgs mass and coupling constant.

The primary focus here lies on infrared divergence. The combined use of RBFM with the Coleman-Weinberg method proves to be highly beneficial in elucidating the intricate points involved in resolving the IR divergences. As employed in the previous section, we utilize the massive Higgs propagator alongside the massless propagators for the gauge and fermion fields. In this section, these fields are ultimately perceived as having temperature-dependent effective masses. Nonetheless, the starting points of all these seemingly distinct treatments stem from RBFM. The apparent differences in treatments merely reflect varying levels of explicitness in the analysis, depending on whether or not the Higgs mass term is explicitly separated out from the Higgs background field $\vf$.

In section 3.1, we undertake the flat spacetime constant-T analysis, detailing the thermal resummation in the gauge and fermion sectors. The expressions of the thermal masses of the gauge and fermionic fields introduced thereby are discussed in section 3.2. We observe in section 3.3 that the main cause for the Linde problem, the absence of the magnetic thermal mass, can be circumvented. Building upon the device proposed in previous work \cite{Park:2021ohu,Park:2021vro}, designed to handle loops in a general curved background at zero-T, we find that it can adeptly address the time-dependent temperature complications as well. In section 3.4, we incorporate the graviton sector and note that the loop analysis essentially reduces to the constant-T flat spacetime case. This implies that the insights gleaned from the analysis in section 3.1 extend beyond their immediate application, as even for a FLRW background, one can leverage the results obtained there. A similar relation for a zero-T curved background was highlighted in \cite{Park:2021ohu}. The results obtained in the present section harbor potentially intriguing implications for the CC problem, effective coupling constants, and Hubble tension, which are pondered in section 4.

\subsection{Finite-T shifts in Higgs mass and coupling \la{thmcr}}

In this section, we employ the perturbative scheme of high-temperature expansion to investigate the finite-temperature behaviors of the Higgs mass and coupling.\footnote{In a more general context, beyond one-loop calculations, it becomes necessary to account for the finite T-dependent terms of the kinetic term as well. Incorporating these terms, particularly when computing propagators, effectively results in a form of resummation. The inclusion of such terms becomes crucial depending on the specific problem under investigation, as they can have significant implications for the overall behavior of the system. Therefore, considering these T-dependent finite terms becomes essential for a comprehensive understanding of the system's dynamics, especially in higher-loop calculations.} Despite the presence of temperature, which breaks Lorentz symmetry (and diffeomorphism symmetry in a curved background), the counter-terms for UV divergences retain the same covariant forms as in the zero-T case, remaining T-independent. This simplifies several aspects of the analysis. While one should ideally consider all diagrams analyzed in the zero-T setup, the renormalization of the Higgs mass and coupling can be addressed without entirely repeating the gauge and fermion sector analyses due to this property.\footnote{An intuitive but non-rigorous way to understand how zero-T renormalization handles UV divergences is by considering a low-T expansion. Since the divergence part is independent of temperature, the renormalization procedure at zero temperature should remain unchanged. For the high-T case, one can imagine summing up the low-T series into a closed form. This closed form can then be extrapolated to the high-T regime, implying that the renormalization procedure for high-T scenarios should be identical to that of the low-T case.

However, for a more rigorous discussion and comprehensive understanding, one should refer to the literature, such as \cite{Laine:2016hma}, which provides detailed analysis and explanations of these concepts.}

For the shifts in the Higgs mass and coupling, we reanalyze the diagrams in Fig. \ref{fig:mren} and \ref{lambdaren} in the finite-T setup. Additionally, we need to consider Fig. \ref{gaugewfr} (c) to tackle the issue of infrared divergence, which necessitates a separate treatment. We revisit the loop analysis, this time focusing on the IR divergences associated with finite-T effects. These divergences can be circumvented through thermal resummation, which is necessary even though we restrict the analysis to one-loop. When applied to the Higgs sector, thermal resummation results in a thermal Higgs mass proportional to $T^2$. Consequently, gauge bosons and fermions will develop mass terms of the same order, leading to several significant implications.

Firstly, the thermal masses enable the resolution of the infrared divergence problem encountered in the computation of some diagrams in Fig. \ref{lambdaren}. Furthermore, the implications for the Linde problem are noteworthy. The cause of the Linde problem can, at least in part, be attributed to the disregard of quark masses compared to the temperature. (See footnote \ref{Lmaincause} for related remarks.) In our dynamic temperature-driven mass renormalization scheme, quark masses cannot be disregarded after resummation since they are of the order of the temperature. Applied to QCD, this implies that both the electric and magnetic components of gluons develop thermal mass terms, thus circumventing the main cause.

\vspace{.2in}
As before, one needs to compute the four diagrams in Fig. \ref{fig:mren}. The diagrams can be evaluated based on the following well known results (see, e.g., \cite{Laine:2016hma}): for a bosonic loop
\bea
J(m,t)&\equiv& \;\; \mathclap{\displaystyle\int}\mathclap{\textstyle\sum}\;\;\;\Big[\fr12\ln(K^2+m^2)-const]=-\fr{\pi^2 }{90}\,T^4 +\fr{m^2 }{24}\,T^2-\fr{1}{12 \pi}m^3 \,T   \nn\\
&&-\fr{ \m^{-2\e}}{2(4\pi)^2}\;m^4 \Big[\ln\Big(\fr{\bar{\m} e^{\g}}{4\pi T}\Big)+\fr1{2\e}\Big]+ \fr{\z(3)}{3(4\pi)^4}\frac{m^6}{T^2}+{\cal O}\Big(\frac{m^8}{T^4}\Big)+{\cal O}(\e) \nn\\
I(m,T)&\equiv& \;\;\mathclap{\displaystyle\int}\mathclap{\textstyle\sum}\;\;\;\fr{1}{K^2+m^2}
 =\fr{T^2}{12}-\fr{mT}{4\pi} -\fr{2 \m^{-2\e}}{(4\pi)^2}\;m^2 \Big[\ln\Big(\fr{\bar{\m} e^{\g}}{4\pi T}\Big)+\fr1{2\e}\Big] \nn\\
 &&\qquad +\fr{2\z(3)}{(4\pi)^4}\frac{ m^4}{T^2}+{\cal O}\Big(\frac{m^6}{T^4}\Big)+{\cal O}(\e)
\label{IJ}
\eea
where $\ln \mb^2 \equiv \ln(4\pi\m^2) -\g$; $\g$ is an Euler's constant. It is also useful to record the expression of $I$ prior to the large-$T$ expansion:
\bea
&& \hspace{.7in} I(m,T)=T\sum_{n= 0} \int \fr{d^d {\bf k}}{(2\pi)^d} \fr{1}{\w_n^2+{\bf k}^2+m^2} \\
&=& \fr{\G(-\fr12+\e)}{(4 \pi)^{\fr32 -\e}} T \,m^{1-2\e}
+\fr{1}{2\; \pi^{2-\frac{d}{2}} T^{1-d}} \sum_{l=0}^\infty \z(2l+2-d)\,\fr{\G(l+1-\fr{d}2)}{\G(l+1)} \fr{(-1)^l m^{2l}}{(2\pi T)^{2l}}. \nn
\eea
Note that in the series term above, the $(m,T)$-dependence comes only in the form $\fr{m^{2l}}{T^{1-d+2 l}}$. For a fermionic loop one instead gets
\bea
\tilde{J}(m.T)\equiv J_0(m)+\tilde{J}_T(m)\quad,\quad \tilde{I}(m.T)\equiv I_0(m)+\tilde{I}_T(m)
\eea
where
\bea
J_0(m) &=&-\fr{m^4 \m^{-2\e}}{64 \pi^2}\Big[\fr1\e+\ln\fr{\mb^2}{m^2}+\fr32+{\cal O}(\e) \Big]\nn\\
I_0(m) &=&-\fr{m^2 \m^{-2\e}}{16 \pi^2}\Big[\fr1\e+\ln\fr{\mb^2}{m^2}+1+{\cal O}(\e) \Big]
\eea
and
\bea
\tilde{J}_T(m)&=& \fr78\fr{\pi^2 }{90}\,T^4 -\fr{m^2 }{48}\,T^2-\fr{ 1}{2(4\pi)^2}\;m^4 \Big[\ln\Big(\fr{m e^{\g}}{\pi T}\Big)-\fr34\Big]+ \fr{7\z(3)}{3(4\pi)^4}\frac{m^6}{T^2}+\cdots \nn\\
\tilde{I}_T(m)&=&- \fr{T^2}{24}-\fr{2 }{(4\pi)^2}\;m^2 \Big[\ln\Big(\fr{m e^{\g}}{\pi T}\Big)-\fr1{2}\Big]+\fr{14\,\z(3)}{(4\pi)^4}\frac{ m^4}{T^2}+\cdots.
\eea
We can illustrate the finite-T computation with several diagrams. Let us illustrate it with the scalar-loop diagram shown in Fig. \ref{fig:mren} (b). This diagram can be computed by applying a finite-T modification of Eq. (\ref{zeroT2ptsl}): replacing $\int d^4k$ by $\;\mathclap{\displaystyle\int}\mathclap{\textstyle\sum}\;\;$, one gets, on account of (\ref{IJ}), 
\bea
\begin{fmffile}{quartichiggssmall}
\!\!\Scale[0.25]{
	\begin{gathered}
		\begin{fmfgraph*}(200,150)
			\fmfleft{i}
			\fmfright{o}
			\fmftop{m}
			\fmf{dashes,tension=1}{i,v1}
			\fmf{dashes,tension=1}{v1,o}
			\fmf{dashes,left,tension=0}{v1,m,v1}
	       \end{fmfgraph*}
	\end{gathered}
	}
\end{fmffile}
= -\l\Big(\int d^4x \, \vF^\dagger \vF\Big) \;\;\;\mathclap{\displaystyle\int}\mathclap{\textstyle\sum}\;\;\;\fr{1}{K^2+m_h^2}
= -\l  \Big( \fr{T^2}{12}-\fr1{4\pi}m_h T+\cdots \Big)\int_x \vF^\dagger \vF.
\eea
The rest of the diagrams can be similarly computed:
\bea
\begin{fmffile}{gaugescalarsmall}
\!\!\Scale[0.13]{
\begin{gathered}
		\begin{fmfgraph*}(300,250)
			\fmfleft{i}
			\fmfright{o}
			\fmfbottom{b} 
			\fmf{dashes,tension=1.3}{i,v1}
			\fmf{dashes}{v1,v2}
			\fmf{dashes,tension=1.3}{v2,o}
			\fmf{gluon,right,tension=-.5}{v2,v1}
			\fmffreeze
	\end{fmfgraph*}
	\end{gathered}
	}
\end{fmffile} \;\;
&=&\fr14   (C_2g^2+g'^2) \Big( \fr{T^2}{12}-\fr1{4\pi}m_h T+\cdots \Big)\int_x \vF^\dagger \vF  \nn\\
\begin{fmffile}{s2ptgaugeloop}
\!\!\Scale[0.26]{
     \begin{gathered}
		\begin{fmfgraph*}(200,150)
			\fmfleft{i}
			\fmfright{o}
			\fmftop{m}
			\fmf{dashes,tension=1}{i,v1}
			\fmf{dashes,tension=1}{v1,o}
			\fmf{gluon,left,tension=0}{v1,m,v1}
	\end{fmfgraph*}
	\end{gathered}
	}
\end{fmffile}  &=&- (C_2g^2+g'^2) \Big( \fr{T^2}{12}+\cdots \Big)\int_x \vF^\dagger \vF \nn\\
\begin{fmffile}{pure2small}
	\!\!\Scale[0.24]{
	\begin{gathered}
		\begin{fmfgraph*}(140,80)
		\fmfleft{i} \fmfright{o}
		\fmf{dashes,tension=5}{i,v1}
		\fmf{dashes,tension=5}{v2,o}
		\fmf{plain,left,tension=1}{v1,v2,v1}
		\end{fmfgraph*}
  \end{gathered}
	}
\end{fmffile} \;\;
&=& 6y_t^2\, \Big(- \fr{T^2}{24}+\cdots \Big)\int_x \vF^\dagger \vF. 
\eea
The first and third diagrams also contribute to kinetic terms. These terms are non-gauge-covariant, yet the breaking effects are subleading in $T$. Summing up, one gets, at the leading order,
\bea
&&\int_x\;\Big\{\Big[- \l+\fr1{4}   (C_2g^2+g'^2)
-   (C_2g^2+g'^2)+6y_t^2 \Big]\fr{T^2}{12}+\cdots\Big\} \vF^\dagger \vF \nn\\
&\simeq&\int_x\;\fr1{12}\Big[- \l-\fr3{4}   (3g^2+g'^2)
-3y_t^2 \Big]{T^2}\;\vF^\dagger  \vF
\la{fthmass}
\eea
which confirms the corresponding part of \cite{Arnold:1992rz}. As for the Higgs coupling renormalization, the zero-T results can again be utilized. However, unlike the mass term, one encounters a complication not present in the zero-T case: the infrared divergence. We illustrate it with the gauge- and fermionic-loop diagrams Fig. \ref{lambdaren} (a) and (b). By utilizing \rf{h4ptgloop}, the gauge-loop diagram (a) is given by:
\bea
\begin{fmffile}{higgs4ptgl2}
\hspace{-.05in}\Scale[0.4]{
       \begin{gathered}
		\begin{fmfgraph*}(70,50)
\fmfleft{i1,i2}
\fmfright{o1,o2}
\fmfbottom{bi,b1,b2,b3,b4,bf}
\fmf{dashes,tension=2}{i1,v1,i2}
\fmf{dashes,tension=2}{o1,v2,o2}
\fmf{gluon,left=1}{v1,v2,v1}
\fmffreeze
	\end{fmfgraph*}
	  \end{gathered}
	              }
\end{fmffile}&=& \fr34 g^4 \;\;\mathclap{\displaystyle\int}\mathclap{\textstyle\sum}\;\;_{\{P_1,P_2,P_3,P_4\}}
\vF^\dagger_{P_1} \vF_{P_2} \, \vF^\dagger_{P_3} \vF_{P_4}\;
 {\raisebox{0.05ex}{$\mathchar '26$}\mkern -9mu\delta}(\Sigma P_l)\;
\;\;\mathclap{\displaystyle\int}\mathclap{\textstyle\sum}\;\; \fr{1}{K^2 (K+P_3-P_4)^2} \la{ftgl4pt}
\eea
where ${\raisebox{0.05ex}{$\mathchar '26$}\mkern -9mu\delta}(\Sigma P_l)$ denotes $(2\pi)^4\d(P_1+P_3-P_2-P_4)$. Let us evaluate the sum-integral over $K^\m$. Since we are interested in the potential it is not necessary to keep track of the $P$-dependent terms. (They will yield terms with derivatives in the position space.) Also, they are subleading in the large-temperature expansion. The $K$-integrand of the above can be expanded as
\bea
\mathclap{\displaystyle\int}\mathclap{\textstyle\sum}\;\ \fr{1}{K^2 (K+P_3-P_4)^2} \simeq \;\;\mathclap{\displaystyle\int}\mathclap{\textstyle\sum}\;\ \left(\fr{1}{K^4}-\fr{2(P_3-P_4)\cdot K}{K^6} -\fr{(P_3-P_4)^2 }{K^6}\right). \la{ftgl4ptexp}
\eea
Postponing the analysis of the higher order terms, such as $\sim\fr{1}{K^6}$, let us evaluate the first two terms. The second term in (\ref{ftgl4ptexp}) vanishes due to the odd parity of the integrand. The leading term $\;\;\mathclap{\displaystyle\int}\mathclap{\textstyle\sum} \;\;  \fr{1}{K^4}$ can be evaluated by taking $\fr1{m} \fr{\pa}{\pa m}$ on $I(m,T)$ followed by a limit $m\ra 0$:
\bea
\;\;\mathclap{\displaystyle\int}\mathclap{\textstyle\sum}\;\;\;\fr{1}{(K^2+m^2)^2}
  &=&-\fr1{2 m}\Big(-\fr{T}{4\pi}-\fr{4 \m^{-2\e}}{(4\pi)^2}\;m \Big[\ln\Big(\fr{\bar{\m} e^{\g}}{4\pi T}\Big)+\fr1{2\e}\Big]
  +\fr{8  \z(3)}{(4\pi)^4}\;\fr{m^3}{T^2} +\cdots  \Big) \nn\\
   &=&\fr{T}{8\pi m}+\fr{2 \m^{-2\e}}{(4\pi)^2} \Big[\ln\Big(\fr{\bar{\m} e^{\g}}{4\pi T}\Big)+\fr1{2\e}\Big]
  -\fr{4  \z(3)}{(4\pi)^4}\;\fr{m^2}{T^2} +\cdots.   
\la{Kto4th}
\eea
If we take the mass of the gauge fields to be zero as in the zero-T analysis, one encounters a problem as one takes the $m\ra 0$ limit, the well-known infrared divergence problem. The same problem occurs in the fermionic-loop diagram, Fig. \ref{lambdaren} (b). By utilizing the zero-T result \rf{h4ptfloop}, one gets
\bea
\begin{fmffile}{higgs4pt13}
\Scale[0.4]{
       \begin{gathered}
		\begin{fmfgraph*}(70,50)
			\fmfleft{i1,i2}
\fmfright{o1,o2}
\fmf{dashes}{i1,v1}
\fmf{plain,tension=.5}{v1,v3}
\fmf{dashes}{v3,o1}
\fmf{dashes}{o2,v4}
\fmf{plain,tension=.5}{v4,v2}
\fmf{dashes}{v2,i2}
\fmf{plain,tension=.5}{v1,v2}
\fmf{plain,tension=.5}{v3,v4}
	\end{fmfgraph*}
	     \end{gathered}
	              }
\end{fmffile}
&=& -{y_t^4}(\d^{\m\n}\d^{\r\s}-\d^{\m\r}\d^{\n\s}+\d^{\m\s}\d^{\n\r})
\;\;\;\mathclap{\displaystyle\int}\mathclap{\textstyle\sum}\;\;_{P's}\;\;
\vF_m(P_1)\vF_n(P_2)\vF_m^\dagger(P_3) \vF_n^\dagger(P_4)\; {\raisebox{0.05ex}{$\mathchar '26$}\mkern -9mu\delta}(\Sigma P_l)\nn\\
&&\qquad\;\;\mathclap{\displaystyle{\tilde{\!\!\!\int}}}\mathclap{\textstyle{\!\!\!\sum}}\;\;\;\fr{K_\m K_\n K_\r K_\s}{K^2 (K-P_1)^2(K-P_1+P_4)^2 (K-P_3)^2}
\eea
where ${\raisebox{0.05ex}{$\mathchar '26$}\mkern -9mu\delta}(\Sigma P_l)$ denotes $(2\pi)^4\d(P_1+P_2-P_3-P_4)$.
Let us expand the $K$-integrand:
\bea
&\simeq&-{y_t^4}\;\;\;\mathclap{\displaystyle\int}\mathclap{\textstyle\sum}\;\;_{P's}\;\;
\vF_m(P_1)\vF_n(P_2)\vF_m^\dagger(P_3) \vF_n^\dagger(P_4) \;{\raisebox{0.05ex}{$\mathchar '26$}\mkern -9mu\delta}(\Sigma P_l)\nn\\
&&\,\;\;\mathclap{\displaystyle{\tilde{\!\!\!\int}}}\mathclap{\textstyle{\!\!\!\sum}}\;\;\;\Big[\fr{1}{K^4}+2\fr{(2P_1-P_4+P_3)\cdot K}{K^6}
-\fr{P_1^2+(P_1-P_4)^2+P_3^2}{K^6} +\cdots \Big].
\eea
Here again the leading order is $\fr1{K^4}$, 
\bea
\;\;\mathclap{\displaystyle{\tilde{\!\!\!\int}}}\mathclap{\textstyle{\!\!\!\sum}}\;\;\;\fr{1}{(K^2+m^2)^2}
  = \fr{\m^{-2\e}}{(4\pi)^2}\Big[\fr{1}{\e}+\ln\fr{\mb^2}{m_f^2}+{\cal O}(\e)\Big]+\fr{2}{(4\pi)^2} \ln\frac{m e^\g}{\pi T}-\frac{28 \z(3)}{(4\pi)^4}\frac{m_f^2}{T^2}\cdots.  \nn\\
\la{Kto4thfer}
\eea
Where the tilde above the integral indicates the fermionic sum-integral. As with the bosonic case, one encounters infrared divergence due to the lack of the mass term. The divergence can be avoided by conducting thermal resummation. Let us pause and recapitulate. Since we are interested in one-loop, it is not necessary, though useful anyway, to consider thermal resummation when it comes to ultraviolet divergence. As the example above reveals, however, it is not the case for infrared divergence: the finite-T effects introduce IR divergence, which in turn is solved by (partially\footnote{It is partially two-loop since we are only considering the two-loop coming from the resummation.} two-loop) finite-T resummation to be spelled out. We will shortly discuss how the mass terms of gauge and fermion fields appear, but for now let us temporarily denote the temperature-dependent gauge and fermion mass parameters by
\be
(m_g, m_f),
\ee 
respectively, and proceed. The results obtained above now take the form:
\bea
\begin{fmffile}{higgs4ptgl2}
\hspace{-.05in}\Scale[0.4]{
       \begin{gathered}
		\begin{fmfgraph*}(70,50)
\fmfleft{i1,i2}
\fmfright{o1,o2}
\fmfbottom{bi,b1,b2,b3,b4,bf}
\fmf{dashes,tension=2}{i1,v1,i2}
\fmf{dashes,tension=2}{o1,v2,o2}
\fmf{gluon,left=1}{v1,v2,v1}
\fmffreeze
	\end{fmfgraph*}
	  \end{gathered}
	              }
\end{fmffile}
&=&\fr34 g^4\Big(\;\;\mathclap{\displaystyle\int}\mathclap{\textstyle\sum}\;\;  \fr{1}{(K^2+m_g^2)^2} +\cdots\Big) \int_x \vF^\dagger \vF \nn\\
&=& \fr34 g^4 \Big(-\fr1{2 m_g}\Big)\Big(-\fr{T}{4\pi}-\fr{4 \m^{-2\e}}{(4\pi)^2}\;m_g \Big[\ln\Big(\fr{\bar{\m} e^{\g}}{4\pi T}\Big)+\fr1{2\e}\Big]
  +\fr{8  \z(3)}{(4\pi)^4}\;\fr{m_g^3}{T^2} +\cdots  \Big)\int_x \vF^\dagger \vF  \nn\\
&=&\fr34 g^4 \left\{\fr{1}{8\pi}\frac{T}{m_g}+\fr{2 \m^{-2\e}}{(4\pi)^2} \Big[\ln\Big(\fr{\bar{\m} e^{\g}}{4\pi T}\Big)+\fr1{2\e}\Big]
  -\fr{4  \z(3)}{(4\pi)^4}\;\fr{m_g^2}{T^2} +\cdots  \right\}\int_x \vF^\dagger \vF  
\eea
\bea
\begin{fmffile}{higgs4pt13}
\Scale[0.4]{
       \begin{gathered}
		\begin{fmfgraph*}(70,50)
			\fmfleft{i1,i2}
\fmfright{o1,o2}
\fmf{dashes}{i1,v1}
\fmf{plain,tension=.5}{v1,v3}
\fmf{dashes}{v3,o1}
\fmf{dashes}{o2,v4}
\fmf{plain,tension=.5}{v4,v2}
\fmf{dashes}{v2,i2}
\fmf{plain,tension=.5}{v1,v2}
\fmf{plain,tension=.5}{v3,v4}
	\end{fmfgraph*}
	     \end{gathered}
	              }
\end{fmffile}
&=&-y_t^4 \Big(\;\;\;\mathclap{\displaystyle{\tilde{\!\!\!\int}}}\mathclap{\textstyle{\!\!\!\sum}}\;\; \fr{1}{(K^2+m_f^2)^2} +\cdots\Big) \int_x \vF^\dagger \vF \\
&=& -y_t^4 \left\{ \fr{\m^{-2\e}}{(4\pi)^2}\Big[\fr{1}{\e}+\ln\fr{\mb^2}{m_f^2}+{\cal O}(\e)\Big]+\fr{2}{(4\pi)^2} \ln\frac{m_f e^\g}{\pi T}-\frac{28 \z(3)}{(4\pi)^4}\frac{m_f^2}{T^2}\cdots  \right\} \int_x \vF^\dagger \vF.  \nn
\eea
The remaining diagrams are more involved but can be similarly computed by utilizing
\bea
\fr{x}{(x + a)^2 (x + b)}&=& \fr{a}{(a - b) (x+a)^2} +\fr{b}{(a - b)^2 (x+a)} -\fr{b}{(a - b)^2 (x+b)} \\
\fr{x^2}{(x + a)^2 (x + b)^2}&=& \fr{a^2}{(a - b)^2 (x+a)^2}+\fr{b^2}{(a - b)^2 (x+b)^2} +\fr{2 a b}{(a - b)^3 ( x+a)} -\fr{2 a b}{(a - b)^3 (x+b)}. \nn
\eea 
Considering only the non-derivative terms, they are, to the leading order in temperature, given by
\bea
\begin{fmffile}{higgs4ptsl2}
\hspace{-.05in}\Scale[0.4]{
       \begin{gathered}
		\begin{fmfgraph*}(70,50)
\fmfleft{i1,i2}
\fmfright{o1,o2}
\fmfbottom{bi,b1,b2,b3,b4,bf}
\fmf{dashes,tension=2}{i1,v1,i2}
\fmf{dashes,tension=2}{o1,v2,o2}
\fmf{dashes,left=1}{v1,v2,v1}
\fmffreeze
	\end{fmfgraph*}
	  \end{gathered}
	              }
\end{fmffile}
&\simeq&\fr{\l^2}{3} \int_x (\varPhi^\dagger \varPhi)^2
\;\;\;\mathclap{\displaystyle\int}\mathclap{\textstyle\sum}\;\;\;\fr{1}{(K^2+m_h^2)^2} 
    \label{scloop4pt}   \nn\\
&=& -\fr1{2 m_h}\Big(-\fr{T}{4\pi}-\fr{4 \m^{-2\e}}{(4\pi)^2}\;m_h \Big[\ln\Big(\fr{\bar{\m} e^{\g}}{4\pi T}\Big)+\fr1{2\e}\Big]
  +\fr{8  \z(3)}{(4\pi)^4}\;\fr{m_h^3}{T^2} +\cdots  \Big)\fr{\l^2}{3} \int_x (\varPhi^\dagger \varPhi)^2 \nn\\
\qquad \begin{fmffile}{higgs4ptggs2}
\hspace{-.05in}\Scale[0.4]{
       \begin{gathered}
		\begin{fmfgraph*}(70,50)
\fmfleft{i1,i2}
\fmfright{o1,o2}
\fmf{dashes,tension=2.2}{i1,g1}
\fmf{dashes,tension=2.2}{g2,o1}
\fmf{dashes,tension=3.2}{i2,v1,o2}
\fmf{gluon,tension=1.2}{v1,g1}
\fmf{gluon,tension=1.2}{v1,g2}
\fmf{dashes,tension=0.5}{g1,g2}
	\end{fmfgraph*}
	  \end{gathered}
	              }
\end{fmffile}
&\simeq& -\fr{5}{8}g^4 \int_x (\varPhi^\dagger \varPhi)^2
\;\;\;\mathclap{\displaystyle\int}\mathclap{\textstyle\sum}\;\;\;\fr{K^2}{(K^2+m_g^2)(K^2+m_g^2)(K^2+m_h^2)}  \nn\\
&& \simeq-\fr{5}{8}g^4 \int_x (\varPhi^\dagger \varPhi)^2
\left\{\Big[\fr{T}{8\pi m_g}+ \fr{2 \m^{-2\e}}{(4\pi)^2}\Big( \fr1{2\e}+\ln \fr{\bar{\m}e^{\g_E}}{4 \pi T}\Big)
-\fr{4 m_g^2 \z(3)}{(4\pi)^4 T^2}\Big] \fr{-m_g^2}{m_h^2-m_g^2} \right.\nn\\
&&\left. +\Big[-\fr{(m_g-m_h)T}{4 \pi}-2 \fr{(m_g^2-m_h^2)\m^{-2\e}}{(4\pi)^2}\Big( \fr1{2\e}+\ln \fr{\bar{\m}e^{\g_E}}{4 \pi T}\Big) +2\z(3) \fr{(m_g^4-m_h^4)}{(4\pi)^4 T^2}\Big] \fr{m_h^2}{(m_h^2-m_g^2)^2} 
\right\} \nn\\
\begin{fmffile}{higgs4ptgss2}
\hspace{-.05in}\Scale[0.4]{
       \begin{gathered}
		\begin{fmfgraph*}(70,50)
\fmfleft{i1,i2}
\fmfright{o1,o2}
\fmf{dashes,tension=2.2}{i1,g1}
\fmf{dashes,tension=2.2}{g2,o1}
\fmf{dashes,tension=3.2}{i2,v1,o2}
\fmf{dashes,tension=1.2}{v1,g1}
\fmf{dashes,tension=1.2}{v1,g2}
\fmf{gluon,tension=0.5}{g1,g2}
	\end{fmfgraph*}
	  \end{gathered}
	              }
\end{fmffile}
&\simeq& -\fr{3}{48}\l g^2 \int_x (\varPhi^\dagger \varPhi)^2
\;\;\;\mathclap{\displaystyle\int}\mathclap{\textstyle\sum}\;\;\;\fr{K^2}{(K^2+m_g^2)(K^2+m_h^2)(K^2+m_h^2)} \nn\\
&& \simeq -\fr{3}{48}\l g^2 \int_x (\varPhi^\dagger \varPhi)^2
\left\{ \Big[\fr{T}{8\pi m_h}+ \fr{2 \m^{-2\e}}{(4\pi)^2}\Big( \fr1{2\e}+\ln \fr{\bar{\m}e^{\g_E}}{4 \pi T}\Big)
-\fr{4 m_h^2 \z(3)}{(4\pi)^4 T^2}\Big] \fr{-m_h^2}{m_g^2-m_h^2}  \right. \nn\\
&&\left. +\Big[-\fr{(m_h-m_g)T}{4 \pi}-2 \fr{(m_h^2-m_g^2)\m^{-2\e}}{(4\pi)^2}\Big( \fr1{2\e}+\ln \fr{\bar{\m}e^{\g_E}}{4 \pi T}\Big) +2\z(3) \fr{(m_h^4-m_g^4)}{(4\pi^4)T^2}\Big] \fr{m_g^2}{(m_g^2-m_h^2)^2} 
\right\} \nn\\
\begin{fmffile}{higgs4ptggss2}
\hspace{-.05in}\Scale[0.4]{
       \begin{gathered}
		\begin{fmfgraph*}(70,50)
\fmfleft{i1,i2}
\fmfright{o1,o2}
\fmf{dashes}{i1,v1}
\fmf{gluon,tension=.5}{v1,v3}
\fmf{dashes}{v3,o1}
\fmf{dashes}{o2,v4}
\fmf{gluon,tension=.5}{v4,v2}
\fmf{dashes}{v2,i2}
\fmf{dashes,tension=.7}{v1,v2}
\fmf{dashes,tension=.7}{v3,v4}
	\end{fmfgraph*}
	  \end{gathered}
	              }
\end{fmffile}
&\simeq&\fr{g^4}{16} \int_x (\varPhi^\dagger \varPhi)^2
\;\;\;\mathclap{\displaystyle\int}\mathclap{\textstyle\sum}\;\;\;\fr{K^4}{(K^2+m_h^2)(K^2+m_h^2)(K^2+m_g^2)(K^2+m_g^2)} \nn\\
&& \simeq\fr{g^4}{16} \int_x (\varPhi^\dagger \varPhi)^2 \left\{\fr{m_h^4}{(m_h^2-m_g^2)^2}\Big[ \fr{T}{8\pi m_h} +\fr{2 \m^{-2\e}}{(4\pi)^2}\Big( \fr1{2\e}+\ln \fr{\bar{\m}e^{\g_E}}{4 \pi T}\Big)  -\fr{4 m_h^2}{(4\pi)^4} \fr{\z(3)}{T^2} \Big]\right. \nn\\
&&\left. +\fr{m_g^4}{(m_h^2-m_g^2)^2}\Big[ \fr{T}{8\pi m_g} +\fr{2 \m^{-2\e}}{(4\pi)^2}\Big( \fr1{2\e}+\ln \fr{\bar{\m}e^{\g_E}}{4 \pi T}\Big)  -\fr{4 m_g^2}{(4\pi)^4} \fr{\z(3)}{T^2} \Big] \right.   \\
&&\left. +\fr{2 m_h^2 m_g^2}{(m_h^2-m_g^2)^3}
\Big[ -\fr{(m_h-m_g)T}{4 \pi} -\fr{2(m_h^2-m_g^2) \m^{-2\e}}{(4\pi)^2}\Big( \fr1{2\e}+\ln \fr{\bar{\m}e^{\g_E}}{4 \pi T}\Big)   +\fr{2 (m_h^4-m_g^4)}{(4\pi)^4} \fr{\z(3)}{T^2}\Big]
\right\}. \nn
\eea

\subsection{Determination of $m_g$ and $m_f$}

It has been noted previously that certain 4-point amplitudes encounter infrared divergences when the gauge and fermion masses are taken to vanish. In the zero-T case, such a problem was not encountered when the gauge mass terms were not considered, as seen in the gauge-loop diagram in equation \rf{higgs2ptcoll}. It was unnecessary to explicitly separate out the vacuum expectation value $\phi_v$ from $\vf$, and the renormalization factors $Z$ for the coupling constants and fields of the unbroken action could be employed without any issue. However, for the Higgs sector, it was essential to focus on the proper Higgs mass term and consider the shift. This was because the goal was not the renormalization of the wrong-sign mass term, but rather that of the proper Higgs mass term of the broken-phase action. Consequently, the vacuum expectation value was separated out at the final stage, and the use of the massive Higgs propagator in the intermediate steps can be understood in this context. Similarly, in the finite-T case, a similar step is necessary for the Higgs sector. Additionally, it is crucial to explicitly separate the vacuum expectation value of $\phi$ at the final stage even for the gauge and fermion sector analyses, but for a different reason: to remove the infrared divergences. Despite the apparent differences in approach between the zero-T and finite-T cases, the refined background field method (RBFM) in the spirit of Coleman-Weinberg provides a unified view, as stated in the introduction to the present section.

Now, let's examine how the mass terms $m_g$ and $m_f$ arise from thermal resummation. Specifically, we will consider the gauge and fermionic-loop diagrams in Fig. \ref{lambdaren} (a) and (b), respectively. (The gauge case is slightly different in that the zero-T Higgs loop also gives rise to a mass term as we observed in section 2.1. However, this contribution is subleading and irrelevant.\footnote{For the Linde problem, however, the analogous term from the fermionic loop should be relevant, as gluons do not couple to the Higgs. Therefore, the mass term from the fermion loop becomes the leading term. Additionally, if one neglects $\tilde{\mu}$, compared to the temperature-induced mass, then such terms become relevant.}) For the gauge loop diagram \ref{lambdaren} (a), let's illustrate the idea with the $U(1)$ gauge field $B_\mu$. The mass term arises from the coupling with the Higgs
\be
-\fr{g'^2}4 \vF^\dagger  \vF\; B_\m^2. \la{Bmass}
\ee
Note that the scalar field here is the background field ${\vF}$ but not the quantum field ${\F}$. Once the field $\vF$ develops a vev, that will lead to $m_g$. To obtain the precise expression of $m_g$, let us switch to the real basis and determine the vev of the scalar field. With thermal resummation in the scalar sector, the vacuum expectation value $\phi_v(T)$ is determined by
\bea
\fr{\pa}{\pa \f}\Big[-\fr{\l}{4!}\f^4+\;\fr12 m_\f^2(T)\f^2 \Big]=0 \quad \Rightarrow \quad\f_v(T)=\sqrt{\fr{6 }{\l}}\;\, m_\f(T)
\eea
where we have chosen the positive branch to be specific, and
\bea
m_\f^2(T)\equiv \tilde{\m}^2-\fr1{12}\Big[ \l+\fr3{4}   (3g^2+g'^2) +3y_t^2\Big]{T^2}. 
\eea
By substituting this into \rf{Bmass}, one gets the expression for $m_g$. Considering the temperature that is substantially higher than the EW scale to which the renormalized mass parameter $\tilde{\m}$ is set, $T\gg \tilde{\m}$, the mass parameter $m_\f(T)$ and $\f_v(T)$ will be
\bea
m_\f^2(T) \sim  T^2\quad,\quad   \f_v^2(T) \sim  T^2.
\eea
Let us introduce a parameter $c_g$ defined by
\bea
m_g(T) \equiv c_g\,  T.
\eea
The parameter $c_g$ is a function the coupling constants; in general it will also depend on one's renormalization scheme. For the fermionic diagram Fig. \ref{lambdaren} (b) (we use the top quark sector to be specific), we view the Yukawa vertex
\be
-y_t\,\barq\,\tilde{\vF}\,t_R+ h.c.
\ee
as the mass term, treating $\tilde{\vF}$ as a constant. Again, note that we use the background field $\tilde{\vF}$ but not the quantum field $\tilde{\F}$. In terms of the complex scalar, the fermionic mass is given by $y_t \tilde{\vF}$. At the final stage one will have to switch to the real basis. Then $\f_v(T)$ emerges through $\tilde{\vF}$; the fermionic mass term is $\fr{1}{\sqrt{2}} y_t \f_v(T)$. We introduce a parameter $c_f$ similarly to the gauge case:
\bea
m_f(T) \equiv c_f\,  T.
\eea
By the same token one can introduce a paramter $c_h$:
\bea
m_h(T) \equiv c_h\,  T.
\eea   
With these definitions, the 4-pt functions previously calculated can be rewritten as
\bea
\begin{fmffile}{higgs4ptgl2}
\hspace{-.05in}\Scale[0.4]{
       \begin{gathered}
		\begin{fmfgraph*}(70,50)
\fmfleft{i1,i2}
\fmfright{o1,o2}
\fmfbottom{bi,b1,b2,b3,b4,bf}
\fmf{dashes,tension=2}{i1,v1,i2}
\fmf{dashes,tension=2}{o1,v2,o2}
\fmf{gluon,left=1}{v1,v2,v1}
\fmffreeze
	\end{fmfgraph*}
	  \end{gathered}
	              }
\end{fmffile}
&=& -\fr38 g^4\Big(-\fr{1}{4\pi c_g}-\fr{4 \m^{-2\e}}{(4\pi)^2}\; \Big[\ln\Big(\fr{\bar{\m} e^{\g}}{4\pi T}\Big)+\fr1{2\e}\Big]
  +\fr{8  \z(3)}{(4\pi)^4}\;{c_g^2} +\cdots  \Big)\int_x \vF^\dagger \vF  \nn\\
\begin{fmffile}{higgs4pt13}
\Scale[0.4]{
       \begin{gathered}
		\begin{fmfgraph*}(70,50)
			\fmfleft{i1,i2}
\fmfright{o1,o2}
\fmf{dashes}{i1,v1}
\fmf{plain,tension=.5}{v1,v3}
\fmf{dashes}{v3,o1}
\fmf{dashes}{o2,v4}
\fmf{plain,tension=.5}{v4,v2}
\fmf{dashes}{v2,i2}
\fmf{plain,tension=.5}{v1,v2}
\fmf{plain,tension=.5}{v3,v4}
	\end{fmfgraph*}
	     \end{gathered}
	              }
\end{fmffile}
&=& -y_t^4 \left\{ \fr{\m^{-2\e}}{(4\pi)^2}\Big[\fr{1}{\e}+\ln\fr{\mb^2}{c_f^2 T^2}+{\cal O}(\e)\Big]+\fr{2}{(4\pi)^2}\ln\frac{c_f e^\g}{\pi }-\frac{28 \z(3)}{(4\pi)^4}c_f^2\cdots  \right\} \int_x \vF^\dagger \vF  \nn\\
\begin{fmffile}{higgs4ptsl2}
\hspace{-.05in}\Scale[0.4]{
       \begin{gathered}
		\begin{fmfgraph*}(70,50)
\fmfleft{i1,i2}
\fmfright{o1,o2}
\fmfbottom{bi,b1,b2,b3,b4,bf}
\fmf{dashes,tension=2}{i1,v1,i2}
\fmf{dashes,tension=2}{o1,v2,o2}
\fmf{dashes,left=1}{v1,v2,v1}
\fmffreeze
	\end{fmfgraph*}
	  \end{gathered}
	              }
\end{fmffile}
&=& -\fr1{2}\Big(-\fr{1}{4\pi c_h}-\fr{4 \m^{-2\e}}{(4\pi)^2}\; \Big[\ln\Big(\fr{\bar{\m} e^{\g}}{4\pi T}\Big)+\fr1{2\e}\Big]
  +\fr{8  \z(3)}{(4\pi)^4}\;{c_h^2} +\cdots  \Big)\fr{\l^2}{3} \int_x (\varPhi^\dagger \varPhi)^2 \nn\\
\qquad \begin{fmffile}{higgs4ptggs2}
\hspace{-.05in}\Scale[0.4]{
       \begin{gathered}
		\begin{fmfgraph*}(70,50)
\fmfleft{i1,i2}
\fmfright{o1,o2}
\fmf{dashes,tension=2.2}{i1,g1}
\fmf{dashes,tension=2.2}{g2,o1}
\fmf{dashes,tension=3.2}{i2,v1,o2}
\fmf{gluon,tension=1.2}{v1,g1}
\fmf{gluon,tension=1.2}{v1,g2}
\fmf{dashes,tension=0.5}{g1,g2}
	\end{fmfgraph*}
	  \end{gathered}
	              }
\end{fmffile}
&& \simeq-\fr{5}{8}g^4 \int_x (\varPhi^\dagger \varPhi)^2
\left\{\Big[\fr{1}{8\pi c_g}+ \fr{2 \m^{-2\e}}{(4\pi)^2}\Big( \fr1{2\e}+\ln \fr{\bar{\m}e^{\g_E}}{4 \pi T}\Big)
-\fr{4  \z(3)}{(4\pi)^4 } c_g^2\Big] \fr{-c_g^2}{c_h^2-c_g^2}\right. \nn\\
&&\left. +
\Big[-\fr{(c_g-c_h)}{4 \pi}-2 \fr{(c_g^2-c_h^2)\m^{-2\e}}{(4\pi)^2}\Big( \fr1{2\e}+\ln \fr{\bar{\m}e^{\g_E}}{4 \pi T}\Big) 
 +2\z(3) \fr{(c_g^4-c_h^4)}{(4\pi)^4 }\Big] \fr{c_h^2}{(c_h^2-c_g^2)^2} 
\right\} \nn\\
\begin{fmffile}{higgs4ptgss2}
\hspace{-.05in}\Scale[0.4]{
       \begin{gathered}
		\begin{fmfgraph*}(70,50)
\fmfleft{i1,i2}
\fmfright{o1,o2}
\fmf{dashes,tension=2.2}{i1,g1}
\fmf{dashes,tension=2.2}{g2,o1}
\fmf{dashes,tension=3.2}{i2,v1,o2}
\fmf{dashes,tension=1.2}{v1,g1}
\fmf{dashes,tension=1.2}{v1,g2}
\fmf{gluon,tension=0.5}{g1,g2}
	\end{fmfgraph*}
	  \end{gathered}
	              }
\end{fmffile}
&& \simeq -\fr{3}{48}\l g^2 \int_x (\varPhi^\dagger \varPhi)^2
\left\{
\Big[\fr{1}{8\pi c_h}+ \fr{2 \m^{-2\e}}{(4\pi)^2}\Big( \fr1{2\e}+\ln \fr{\bar{\m}e^{\g_E}}{4 \pi T}\Big)
-\fr{4  \z(3)}{(4\pi)^4 } c_h^2\Big] \fr{-c_h^2}{c_g^2-c_h^2} \right. \nn\\
&&\left. +\Big[-\fr{(c_h-c_g)}{4 \pi}-2 \fr{(c_h^2-c_g^2)\m^{-2\e}}{(4\pi)^2}\Big( \fr1{2\e}+\ln \fr{\bar{\m}e^{\g_E}}{4 \pi T}\Big) +2\z(3) \fr{(c_h^4-c_g^4)}{(4\pi^4)}\Big] \fr{c_g^2}{(c_g^2-c_h^2)^2} 
\right\} \nn\\
\begin{fmffile}{higgs4ptggss2}
\hspace{-.05in}\Scale[0.4]{
       \begin{gathered}
		\begin{fmfgraph*}(70,50)
\fmfleft{i1,i2}
\fmfright{o1,o2}
\fmf{dashes}{i1,v1}
\fmf{gluon,tension=.5}{v1,v3}
\fmf{dashes}{v3,o1}
\fmf{dashes}{o2,v4}
\fmf{gluon,tension=.5}{v4,v2}
\fmf{dashes}{v2,i2}
\fmf{dashes,tension=.7}{v1,v2}
\fmf{dashes,tension=.7}{v3,v4}
	\end{fmfgraph*}
	  \end{gathered}
	              }
\end{fmffile}
&& \simeq\fr{g^4}{16} \int_x (\varPhi^\dagger \varPhi)^2 \left\{\fr{c_h^4}{(c_h^2-c_g^2)^2}\Big[ \fr{1}{8\pi c_h} +\fr{2 \m^{-2\e}}{(4\pi)^2}\Big( \fr1{2\e}+\ln \fr{\bar{\m}e^{\g_E}}{4 \pi T}\Big)  -\fr{4 \z(3)}{(4\pi)^4}c_h^2 \Big]\right. \nn\\
&&\left. +\fr{c_g^4}{(c_h^2-c_g^2)^2}\Big[ \fr{1}{8\pi c_g} +\fr{2 \m^{-2\e}}{(4\pi)^2}\Big( \fr1{2\e}+\ln \fr{\bar{\m}e^{\g_E}}{4 \pi T}\Big)  -\fr{4 \z(3)}{(4\pi)^4}  c_g^2\Big] \right. \nn\\  
&&\left. +\fr{2 c_h^2 c_g^2}{(c_h^2-c_g^2)^3}
\Big[ -\fr{(c_h-c_g)}{4 \pi} -\fr{2(c_h^2-c_g^2) \m^{-2\e}}{(4\pi)^2}\Big( \fr1{2\e}+\ln \fr{\bar{\m}e^{\g_E}}{4 \pi T}\Big)   +\fr{2 \z(3)}{(4\pi)^4}  (c_h^4-c_g^4)\Big]
\right\}. \nn \\ \la{fig5wrs}     
\eea
In the fermion sectors, particularly for masses much smaller than the top quark mass, the values of $c_f$ might be small. A small $c_f$ results in a large logarithmic term in the fermion-loop diagram. It's anticipated that these large logarithmic terms can be managed through renormalization group techniques.


\subsection{On the Linde problem}

While this section may diverge from the main theme of the present work, the thermal resummation approach described above readily provides valuable insights into tackling a longstanding issue in QCD: the Linde problem, also known as the infrared problem, which arises from the absence of the magnetic thermal mass of the gauge fields in QCD.\footnote{
All the quantities influenced by the Linde problem can be accurately determined using lattice techniques, which is highly impressive. The insights offered by our current approach are analytical in nature. The magnetic thermal mass serves as a direct indicator of the non-perturbative dynamics inherent in the problem. I extend my gratitude to M. Laine for the discussion and clarification on this matter.
 \la{Lmaincause}}  Consider eq. \rf{g2ptfl}. Its finite-T counterpart is:
\bea
\begin{fmffile}{gauge2ptfL}
	\!\!\Scale[0.18]{
		\begin{gathered}
		\begin{fmfgraph*}(160,100)
		\fmfleft{i} \fmfright{o}
		\fmf{gluon,tension=6}{i,v1} \fmf{gluon,tension=6}{v2,o}
		\fmf{plain,left,tension=1.3}{v1,v2,v1}
		\end{fmfgraph*}
		\end{gathered}
	}
\end{fmffile}&=&-g^2 \Ct_2 \;\;\;\;\mathclap{\displaystyle{\tilde{\!\!\!\int}}}\mathclap{\textstyle{\!\!\!\sum}}\;\;\;
\fr{{\raisebox{0.05ex}{$\mathchar '26$}\mkern -9mu\delta}(P_1+P_2)}{(K^2+m_f^2)\Big[(K-P_2)^2+m_f^2\Big]}
\Big[K\cdot A^a(P_1) \;(K-P_2)\cdot A^a(P_2)\nn\\
&&+(K-P_2)\cdot A^a(P_1) \;K\cdot A^a(P_2)
-K\cdot (K-P_2)\; A^a(P_1)\cdot A^a(P_2)
\Big] \\
&=&-g^2 \Ct_2 \;\;\;\;\mathclap{\displaystyle{\tilde{\!\!\!\int}}}\mathclap{\textstyle{\!\!\!\sum}}\;\;\;
\fr{K^\m (K-P_2)^\n+(K-P_2)^\m K^\n-K^\r (K-P_2)_\r\; \d_{\m\n}}{(K^2+m_f^2)\Big[(K-P_2)^2+m_f^2\Big]}\;
 A_\m^a A_\n^a  \;{\raisebox{0.05ex}{$\mathchar '26$}\mkern -9mu\delta}(\Sigma P).  \nn
\eea
By focusing on the $P$-independent terms, one gets
\bea
&& \hspace{1.4in}\Rightarrow \quad-g^2 \Ct_2 \;\;\;\;\;\mathclap{\displaystyle{\tilde{\!\!\!\int}}}\mathclap{\textstyle{\!\!\!\sum}}\;\;\;
\fr{2K^\m K^\n-K^2 \d_{\m\n}}{(K^2+m_f^2)^2} \nn\\
&& \hspace{-.3in}= g^2 \Ct_2 \left\{\Big[\d_{\m0}\d_{\n0}+\Big(1-\fr2d\Big)\d_{\m i}\d_{\n i} \Big]\;\;\;\;\mathclap{\displaystyle{\tilde{\!\!\!\int}}}\mathclap{\textstyle{\!\!\!\sum}}\;\;
\fr{K^2 }{(K^2+m_f^2)^2} 
+\Big[-2\d_{\m0}\d_{\n0}+\fr2d\d_{\m i}\d_{\n i} \Big]\;\;\;\;\mathclap{\displaystyle{\tilde{\!\!\!\int}}}\mathclap{\textstyle{\!\!\!\sum}}\;\;
\fr{k_n^2}{(K^2+m_f^2)^2} 
\right\}.  \la{g2ptfloopint} \nn\\
\eea
One can show
\bea
\;\;\;\;\;\mathclap{\displaystyle{\tilde{\!\!\!\int}}}\mathclap{\textstyle{\!\!\!\sum}}\;
\fr{K^2}{(K^2+m_f^2)^2} 
=\tilde{I}(m_f,T)+\frac{m_f}{2}\fr{\pa}{\pa m_f} \tilde{I}(m_f,T).
\eea
Since the mass depends on the temperature, $m_f=m_f(T)$, care is required when taking $\fr{\pa}{\pa T}$ on $\tilde{I}(m,T)$. One can also show 
\bea
\;\;\;\;\;\mathclap{\displaystyle{\tilde{\!\!\!\int}}}\mathclap{\textstyle{\!\!\!\sum}}\;\;
\fr{k_n^2}{(K^2+m_f^2)^2} 
=\fr12 \tilde{I}(m_f,T)
-\fr{T}{2}\fr{\pa}{\pa T} \tilde{I}(m_f,T)
+\fr{T}{4 m_f}\Big(\fr{\pa m_f^2}{\pa T}\Big)\fr{\pa}{\pa m_f} \tilde{I}(m_f,T).
\eea
Substituting these into \rf{g2ptfloopint} one gets
\bea
&&\hspace*{-.5in}-g^2 \Ct_2 \;\;\;\;\mathclap{\displaystyle{\tilde{\!\!\!\int}}}\mathclap{\textstyle{\!\!\!\sum}}\;\;
\fr{2K^\m K^\n-K^2 \d_{\m\n}}{(K^2+m_f^2)^2} \nn\\
&&\hspace*{-.5in}=g^2 \Ct_2 \left\{\d_{\m0}\d_{\n0}\Big[{T}\fr{\pa}{\pa T} \tilde{I}+\fr{m_f}{2}\fr{\pa}{\pa m_f} \tilde{I}
-T\Big(\fr{\pa m_f}{\pa T}\Big)\fr{\pa}{\pa m_f} \tilde{I} \Big]
\right. \nn\\
&&\hspace*{-.5in}+\left. \fr13\d_{\m i}\d_{\n i}\Big(2\tilde{I}-{T}\fr{\pa}{\pa T} \tilde{I}+\fr{m_f}{2}\fr{\pa}{\pa m_f} \tilde{I}
+{T}\Big(\fr{\pa m_f}{\pa T}\Big)\fr{\pa}{\pa m_f} \tilde{I}\Big)  \right\}.
\eea
The explicit expression can be straightforwardly found by substituting $\Ct_2=2,\, \tilde{I},\, \fr{\pa}{\pa T} \tilde{I}$, and $\fr{\pa}{\pa m_f} \tilde{I}$. Unlike the analysis by employing the conventional scheme \cite{Laine:2016hma}, the coefficient of $\d_{\m i}\d_{\n i}$ no longer vanishes, which implies the presence of magnetic thermal mass. Since the magnetic thermal mass serves as a direct indicator of the non-perturbative dynamics, the present result indicates, at least as far as we can tell (see also footnote \ref{Lmaincause}), that the Linde problem can be avoided by a smart choice of the renormalization scheme.

\subsection{A curved background with time-dependent temperature}

In order to achieve a more comprehensive understanding of early Universe physics, particularly in relation to the cosmological constant problem and the Hubble tension, it is imperative to develop a formalism for quantum gravitational thermodynamics. Such a formalism should account for both quantum effects and finite-temperature effects. In this section, we contemplate the methodology for conducting time-dependent finite-temperature analysis within a FLRW background.

While there are circumstances where higher-derivative quantum-correction terms cannot be disregarded \cite{Nurmagambetov:2018het,Nurmagambetov:2020ann}, the present case is likely one where the leading-order action suffices for cosmological purposes. Contributions of gravitons are expected to be small and unlikely to alter the qualitative features observed so far. Nonetheless, it's beneficial to consider how to undertake time-dependent finite-temperature quantum-gravitational perturbation theory to ensure the validity of subsequent discussions.

The FLRW spacetime background introduces nontrivial technical challenges: it is both curved and has a time-dependent temperature. As for the complications of a curved background at zero-T, an effective method for calculating curved-spacetime Feynman diagrams has been devised \cite{Park:2021ohu}. Another major obstacle is how to perform quantum gravitational thermodynamics. While a full solution to these problems won't be pursued here, we propose a prescription that seems natural and readily applicable - which also seems promising for the perspective of a full solution - to relatively simple spacetimes like the FLRW universe. The primary tool is again the general form of the propagator obtained in previous work \cite{Park:2021ohu}. By employing this tool, the analysis is simplified to calculating diagrams in a constant-temperature flat-spacetime background.

Our thermodynamics prescription is based on the well-known equilibrium form of the density operator. Several justifications support using this form here. Firstly, although the Universe was not always in equilibrium, it was so for most of its history. Even when it was not in equilibrium, equilibrium thermodynamics often provides a good approximation. Secondly, it's plausible to extend the definition of temperature (and equilibrium) to a more general notion, such as time- and/or space-dependent temperature. Evidently, the temperature of the FLRW background varies with time. (Position-dependent temperature was also introduced in the literature.) Furthermore, it's possible to have a nontrivial temperature for a wide range of spacetimes, as temperature can be introduced as a Lagrange multiplier in entropy maximization procedures \cite{Sakurai}.

With these considerations, we extend the constant finite-temperature formalism to the time-dependent case by replacing the constant temperature in the Euclidean-time integral with a time-dependent one, scaling as $T(t)\sim \frac{1}{a(t)}$, where $a$ denotes the scale parameter. To illustrate this, let us review the zero-temperature curved-spacetime Feynman diagram analysis by focusing on the metric. We split the metric into solution background and quantum fluctuation components:
\bea
g_{\m\n}\ra h_{\m\n}+{g_s}_{\m\n}
\la{gshift}
\eea
where $g_{s\, \m\n}$ denote the classical solutions and $h_{\m\n}$ the fluctuation fields.\footnote{For the sake of simplicity in our discussion, we will omit the consideration of the background field, which would be the metric analog of $\phi_B$; the discussion can be extended to include it.}
The zero-temperature loop analysis is based on the following two-point function (see \cite{Park:2019amz} for the conventions):
\bea 
<h_{\m\n}(x)h_{\r\s}(y)>&=& {P}_{\m\n\r\s}\, {\D}(x-y)  \la{h2pt}
\eea
where the tensor ${P}_{\m\n\r\s}$ is given, in de Donder gauge, by \cite{Park:2014noa}
\bea
{P}_{\m\n\r\s} &\equiv& \fr{\bar{\k}^2}2\Big({g_s}_{\m\r}{g_s}_{\n\s}+{g_s}_{\m\s}{g_s}_{\n\r}
- \fr12 {g_s}_{\m\n}{g_s}_{\r\s}\Big);   \la{fpt}
\eea
where $\bar{\k}^2\equiv 2\k^2$; ${\D}(x-y)$ is Green's function for a massless scalar theory\footnote{If one views the CC term as contributing to the graviton mass, ${\D}(x-y)$ should be taken as the propagator of a massive scalar theory.}: 
\bea
{\D}(x-y)=\int \fr{d^4k}{(2\pi)^4}\fr1{\sqrt{-{g_s}(x)}}\; \fr{e^{ik\cdot (x-y)}}{i k_\m k_\n {g_s}^{\m\n}(x)}. \la{fcp}
\eea
The propagator can then be transformed as follows: defining the ``flattened" momentum and coordinates, $(\tilde{k}, \tilde{x})$, as
\bea
\tilde{k}_{\underline{\a}}\equiv {e_s}_{\underline{\a}}^\m\, k_\m\quad,\quad  \tilde{x}^{\underline{\b}} \equiv {e_s}^{\underline{\b}}_\n \,x^\n  \quad,\quad
{e_s}_{\underline{\a}}^\m {e_s}_{\underline{\b}}^\n\, {g_s}_{\m\n}=\eta_{\underline{\a}\underline{\b}}
\eea
where the underlined indices denote the flattened ones, one gets the flattened propagator:
\bea
{\D}(\tilde{x}-\tilde{y})=\int \fr{ d^4\tilde{k}}{(2\pi)^4}\fr{e^{i\tilde{k}_{\underline{\g}} (\tilde{x}-\tilde{y})^{\underline{\g}}}}{i \tilde{k}_{\underline{\a}} \tilde{k}_{\underline{\b}} \h^{\underline{\a}\underline{\b}}}.
\la{ffp}
\eea
With this, the steps for evaluating Feynman diagrams in a curved spacetime become parallel to those for the flat case.

Considering the time-dependent temperature in a FLRW background, although $T(t)$ is not constant, it is covariantly constant. This is because the time-dependence of $T(t)$ comes from $a(t)$, which is a metric component and is thus annihilated by a covariant derivative. In the usual comoving coordinates, the mode associated with the time-direction takes $\omega_n=2\pi T(t) n$. By introducing "flattened" coordinates of the comoving coordinates, the time-dependent-T curved-spacetime analysis parallels that of the constant-T flat-spacetime, with all curvature effects included in the formalism. However, there remains some arbitrariness associated with choosing the finite parts, akin to the usual infinities in quantum field theory. Detailed analysis of loops involving the SM fields and gravitons will demand extensive and intensive efforts and warrants dedicated research.

\section{Implications for cosmology}

With the UV and IR divergences at finite temperature duly addressed, let us look into the cosmological implications of our findings, focusing on the cosmological constant (CC) problem and the Hubble tension. We consider the leading part of the 1PI effective action in the derivative expansion coupled with conventional hydrodynamic or kinetic-theory matter. The hydrodynamic matter field equations, i.e., the conservation equations, remain unchanged. As for the field equations of Standard Model matter, we set all, except the Higgs, of the fields to zero. In both the metric field equation and the SM matter field equations, various coupling constants are replaced by their T-dependent effective counterparts.

The essence of our analysis regarding the finite-temperature effects on the CC problem was previously presented in our works \cite{Park:2021ohu,Park:2021vro}. In those studies, employing an Einstein-scalar system, we demonstrated that the CC problem transitions from an order-60 problem to an order-4 problem. It was anticipated that the inclusion of SM matter fields would further narrow this order gap. We provide a detailed explanation here. Regarding the Hubble tension, there are several factors that deviate the system from conventional treatments. Most notably, the effective coupling constants appearing in the field equations are all temperature-dependent, hence time-dependent.

A series of works on a vacuum model, the so-called running vacuum model \cite{Sola:2007sv,Fritzsch:2012qc,Moreno-Pulido:2020anb,SolaPeracaula:2022hpd,Moreno-Pulido:2022phq,Moreno-Pulido:2023ryo}, have been conducted by J.~Solà Peracaula and collaborators. While differing in specifics, their approach and ours share certain conceptual features under the broader umbrella of dynamical vacuum energy models. In particular, both frameworks employ novel renormalization schemes to address the cosmological constant problem. In our case, the time dependence of the coupling constants emerges from finite-temperature loop effects. By contrast, in the running vacuum model approach, adiabatic regularization is employed and the renormalization scale is identified with the Hubble expansion rate $H$. Given the inherent flexibility in choosing renormalization schemes and conditions, we view these as two complementary avenues. Although the intermediate theoretical treatments differ, it is plausible that the final physical predictions - such as vacuum energy evolution or effective cosmological observables - ultimately converge.\footnote{As stated in the notes added at the end, our recent numerical studies indicate that the leading-order finite-T effects do not ameliorate the Hubble tension.  In contrast, see, e.g., \cite{SolaPeracaula:2021gxi} for a work where a novel renormalization scheme {\em does} ameliorate the tension.}

\subsection{Cosmological constant problem}

In \cite{Park:2021ohu,Park:2021vro}, a CC-problem proposal of a novel renormalization scheme was put forth. One crucial aspect was setting the renormalized mass of the scalar field to the order of the Cosmic Microwave Background temperature. However, there were some loose ends in the analysis. Firstly, the extension of the results from constant temperature QFT analysis to a time-dependent temperature scenario lacked rigorous treatment of the time-dependent finite-T QFT procedure. This gap has been addressed in the previous section. Secondly, there remained a significant factor of $\sim 10^4$ or $10^5$ difference between the observed and theoretical values. As anticipated, this discrepancy should be partly attributed to the fact that the system considered there was an Einstein-scalar system. Here, we confirm that the inclusion of SM particle content further narrows this gap. As mentioned in section 3.4, one can consider a flat background and constant temperature. (The result for the FLRW spacetime is obtained simply by replacing $T\ra \frac{T}{a^4}$ at the end.)

In the previous work, the comparison between the observed value and theoretical one was made in the energy ($GeV$) unit. Here, we repeat the comparison, but the comparison will be made in the $[L]^{-4}$ unit for a cross-check. The observed value of the vacuum energy density is
\bea
\r_\L \approx 3.7\times 10^{9} \;eV m^{-3}.
\eea
One can go from $GeV$ to $m$ by the relation (see, e.g., \cite{Gorbunov}) 
\bea
GeV =0.5 \times 10^{14}\;cm^{-1}=0.5 \times 10^{16}\;m^{-1} \la{Rubakovscaling}
\eea
from which one gets
\bea
\r_\L \approx 3.7\times 10^{9} \times 0.5\times 10^7 m^{-4}\approx  10^{16} m^{-4}.
\eea
Let us compare this with the theoretical value based on our proposal. The temperature is 2.7 K and
\bea
1\;K=8.6\times 10^{-14}\;GeV =8.6 \times 10^{-14}\times 0.5\times 10^{16} \;m^{-1}=4\times 10^2 \;m^{-1}.
\eea
Since the temperature-dependent CC term scales as $\sim T^4$,\footnote{The reason for employing a high-T expansion despite the fact that the CMB temperature is considered was explained in \cite{Park:2021ohu}.}   one gets, for the present CC,
\bea
\sim(2.75 K)^4 \approx  10^{12}\;m^{-4},
\eea
thus the gap between the experimental and theoretical values is on the order $10^4$ or so.

As anticipated in \cite{Park:2021ohu}, the SM particle content should further reduce the gap.\footnote{Naively, it appears that the finite-T loop-induced CC is redundant since the radiation is already considered. However, there is the following subtle and interesting fact: particles contribute to vacuum energy through an off-shell virtual channel, whereas radiation arises from their physical or on-shell nature.} When counting the number of fermionic degrees of freedom, there is a delicate point to consider. The total number of fermionic degrees of freedom depends on whether neutrinos are treated as Dirac spinors or Majorana spinors. In this work, we treat them as Dirac spinors. Since our interest lies in the order of magnitude of the CC, our qualitative conclusion remains unchanged even if neutrinos are treated as Majorana spinors.\footnote{For a Majorana spinor, going from Minkowski spacetime to Euclidean spacetime involves doubling of degrees of freedom. An Euclideanization method that avoids this doubling was proposed in the past; for instance, see \cite{vanNieuwenhuizen:1996tv}.} Here's how the counting goes: for a Dirac spinor, which has 4 components, the result of its one-loop contribution to the free energy density is given by 
\bea
-4\;\;\;\mathclap{\displaystyle{\tilde{\!\!\!\int}}}\mathclap{\textstyle{\!\!\!\sum}}\;\;\; \fr12 \ln (K^2+m_f^2)+\mbox{(irrelevant constant)}.
\la{countingJ}
\eea
As we have reviewed in the previous section, 
\bea
\;\;\mathclap{\displaystyle{\tilde{\!\!\!\int}}}\mathclap{\textstyle{\!\!\!\sum}}\;\;\; \fr12 \ln (K^2+m_f^2)=\fr78 \fr{\pi^2}{90}T^4+\cdots.
\eea
For the quarks, the counting is $3\times 3 \times 4\times 2=72$, where the 3's come from the colors and flavors. The factor 4 arises from Eq. \eqref{countingJ}, and the last factor 2 stems from considering both the left-handed and right-handed quarks as "doubles" in the counting. Similarly, when counting the contributions from charged leptons, the total count is $72+12=84$.

Suppose we treat the neutrinos as Dirac spinors. Since there are six of them, the count adds up to 6, resulting in a grand total of $84+12=96$. However, if we treat them as Majorana spinors, the count becomes $84+6=90$.\footnote{Strictly speaking, one would also have to consider the graviton contributions of two physical states, to be precise.} These contributions, with a weight of $\frac{7}{8}$, should be added to the contribution from the 28 bosonic degrees of freedom, each with a weight of 1. With this total, the CC problem is reduced from a $10^{60}$-order problem to a $10^{2}-10^{3}$-order problem.

Now, let's pause and reassess the renormalization of the cosmological constant. The component we've been considering is the one-loop correction, and we've demonstrated that it is two or three orders of magnitude smaller than the observed value. However, it's important to remember that we can introduce a classical piece to the CC. It's not out of the norm but well within the standard practice of renormalization procedure to introduce a classical term, i.e., the renormalized CC, that is two or three orders of magnitude larger than the one-loop quantum corrections. In other words, we no longer face the unnaturalness of dealing with adding and/or subtracting extremely large and small numbers.

\subsection{Effective constants, Hubble tension, and CC non-conservation}

Another potentially significant cosmological implication of our present results concerns the Hubble tension. The conventional Standard Model of cosmology utilizes a classical Einstein system coupled with kinetic-theory matter. However, we propose a more general approach where we consider a finite-temperature system comprising two components: the first component is the leading-order effective action of the Standard Model coupled with gravity, and the second component is the hydrodynamic or kinetic matter system coupled to the first. In finding solutions to the field equations, all of the Standard Model field vacuum expectation values except that of the Higgs field can be set to zero.

Compared with the conventional Standard Model of cosmology, our approach incorporates several additional effects. Most notably, our system includes finite-temperature quantum-field-theoretic effects, whereas in the conventional analysis, only the thermodynamics of the kinetic matter is considered. The consequential main difference in our approach is that all coupling constants, including the cosmological constant (CC) and Newton's constant, are replaced by temperature-dependent effective ones. Recall the renormalization method employed in our work, which is essentially the $\overline{MS}$ scheme. Thus, the renormalization group running of the fundamental coupling constants remains the same as that of the zero-temperature theory. However, the coupling constants appearing in the action or field equations used in cosmology are the effective ones.

Let's examine the CC and Einstein-Hilbert sector to scrutinize the potential implications of the effective coupling constants for Hubble tension. Consider the metric field equation:
\bea
R_{\m\n}-\fr12 g_{\m\n}R+\L_{eff}(t) \;g_{\m\n}=8 \pi \, G_{eff}(t)\; T_{\m\n}
\eea
where $T_{\m\n}$ contains all of the matter sectors. In the conventional analysis, both effective couplings $\L_{eff}$ and $G_{eff}$
are treated as constants. Since they are now time-dependent, it is natural to expect that their presence will modify the best fit cosmological parameters of the CMB power spectrum. Currently a numerical analysis is being conducted \cite{Wui-Park} to see whether the finite-T shift of the cosmological constant could ameliorate the Hubble tension.\footnote{Finite-temperature effects introduce additional density parameters in the Hubble constant, $H(t)$, the first two leading terms of which are denoted as $\frac{\Omega_{\L_2}}{a^4}+\frac{\Omega_{\L_3}}{a^2}$ in \cite{Wui-Park}. The power spectrum sensitively depends on these parameters. See the notes added at the end of the conclusion for the numerical results that recently came out. \la{ns}} A perhaps more profound potential implication of the time-dependent effective coupling constants is for non-conservation of the vacuum energy. Taking a divergence on both sides and substituting
\bea
\nabla^\m (R_{\m\n}-\fr12 g_{\m\n}R)=0
\eea
where the covariant derivative is one associated with quantum-corrected metric, and 
\bea
\nabla^\m T_{\m\n}=0
\eea
(for a more general case, see the conclusion), one gets
\bea
\nabla_\n \L_{eff}(t)= 8 \pi  T_{\m\n} \nabla^\m G_{eff}(t).  \la{CCcons}
\eea
 Note that the (non)constancy of $\L_{eff}$ and $G_{eff}$ is interconnected: if $\L_{eff}$ were a constant, it implies that the right-hand side vanishes (hence the constancy of $G_{eff}$), and this equation can be interpreted as conservation of vacuum energy. However, the non-constancy of $\L_{eff}$, implies that the vacuum energy, unlike the matter or radiation sector, is not conserved. In particular, the effect feeds back to the Newton's constant. We will discuss this further in the conclusion.

\section{Conclusion}

In this work, we have brought forth a unified technical framework provided by RBFM and analyzed the Standard Model coupled to dynamic gravity in a FLRW background. Several technical hurdles have been overcome. Firstly, we addressed the IR divergences by employing thermal resummation not only in the scalar sector but also in the gauge and fermionic sectors. We have highlighted that this procedure allows one to avoid the main cause of the Linde problem. Secondly, progress has been made on how to perform finite-T QFT with time-dependent temperature. This problem may be considered in the context of more general spacetimes: by employing "flat" coordinates, the renormalization analysis becomes analogous to that of a flat spacetime.\footnote{
Relatedly, more work is needed to generalize the concept of equilibrium: temperature can be introduced as a Lagrange multiplier, which need not necessarily be a constant in general. One could argue that the Lagrange multiplier assumes the role of temperature at equilibrium. It appears that temperature can be defined more broadly beyond the standard notion of equilibrium.}

Our method is optimized for computing the 1PI effective action. For illustrations, we computed several sectors of the effective action, including the Higgs sector. The temperature-dependent terms are ubiquitous and call for consideration in cosmology. For cosmological applications, the leading order of the effective action should be sufficient. It should then be coupled to the hydrodynamic or kinetic matter system. It was noted in an $\overline{MS}$ renormalization scheme that the effective coupling constants appearing in the cosmological Lagrangian are all temperature-, thus time-, dependent, although the fundamental coupling constants still have the same renormalization group flow as in the zero-T case.

Two cosmological applications were considered: the CC problem and Hubble tension. By tightening a couple of loose ends in the earlier analysis in \cite{Park:2021ohu}\cite{Park:2021vro}, we have observed that one no longer encounters the need for fine-tuning: the one-loop correction is smaller than the observed value of the CC by a few orders. It is not unnatural to introduce a classical piece of the CC that is a few orders larger than the one-loop result. The appearance of the effective coupling constants should definitely have some impact on analyses, such as the CMB power spectrum. As a preliminary discussion, we have pointed out that the temperature effects have the potential to ameliorate Hubble tension. More profoundly, the "conservation law" given in \rf{CCcons} may well signify certain mysterious aspects of our Universe.

\vspace{.2in}

There are several future directions worth pursuing.
\vspace{.1in}

Perhaps the most urgent direction is to analyze the finite-T effects on cosmological parameters, particularly the Hubble parameter. Currently, work is underway to numerically study these effects \cite{Wui-Park} (see the notes added below for the numerical results that just came out). Additionally, it will be fascinating to explore the intriguing aspect of the temperature effect observed at the end of section 4.2: the constraint between the renormalization of the CC and Newton's constant. In standard cosmology, separate conservation of each matter sector is enforced. Genuine constancy of the CC trivially satisfies this requirement. However, as observed at the end of section 4.2, the time dependence of the CC deviates from this, which, in turn, should have cosmological implications on other matter sectors through gravitational interactions.
One can extend the non-conservation of the CC further in a more general context. In \cite{Park:2018xtt}, it was noted that a Noether current fails to be conserved, meaning it takes on a different form, when a Neumann boundary condition is applied. Similarly, one could extend this observation to the level of equations of motion: relaxing the requirement for separate conservation allows for the consideration of non-conservation in at least some of the matter components.

Another direction, although less urgent, is to gather more examples demonstrating the compatibility of different renormalization schemes. While the finite renormalization-group invariance of physical quantities ensures this compatibility, explicitly demonstrating it would be worthwhile.

\vspace{1in}
\ni {\bf Notes added:}
\vspace{.1in}

\ni Building on the finite-temperature framework developed in this work, we have recently extended our analysis to directly confront observational data within the $\Lambda$CDM paradigm and its finite-temperature modification \cite{Park:2025iog,Hatefi:2025lqe}. Specifically, we investigated the implications of temperature-dependent quantum field theory corrections to the cosmological constant, which lead to the introduction of additional effective density parameters, $\Omega_{\Lambda_2}$ and $\Omega_{\Lambda_3}$, arising from finite-$T$ quantum gravitational effects. Employing the Cosmic Linear Anisotropy Solving System (CLASS) \cite{Blas:2011rf}, we examined the impact of these corrections on the CMB power spectrum and compared the results with Planck 2018 data \cite{Planck:2018vyg}. Through a combination of brute-force parameter scans and modern machine learning techniques, we found that the inclusion of $\Omega_{\Lambda_2}$ and $\Omega_{\Lambda_3}$ enhances the model’s fit to observations, achieving high values of the coefficient of determination, $R^2$, (a higher value of $R^2$ means a better fit) and low mean squared error. Although the results obtained in \cite{Park:2025iog,Hatefi:2025lqe} based on point estimates do not alleviate the Hubble tension, they support the physical viability of finite-temperature corrections in cosmological modeling and open a path toward incorporating higher-order effects and more sophisticated computational tools in future studies.

\newpage
\appendix

\newpage

\end{document}